\documentclass[lettersize,journal]{IEEEtran}
\usepackage{amsmath,amsfonts}
\usepackage{cite}
\usepackage{amsmath,amssymb,amsfonts}
\usepackage{graphicx}
\usepackage{textcomp}
\usepackage{xcolor}
\usepackage[left]{lineno}
\def\BibTeX{{\rm B\kern-.05em{\sc i\kern-.025em b}\kern-.08em
    T\kern-.1667em\lower.7ex\hbox{E}\kern-.125emX}}
\usepackage[colorlinks=true, linkcolor=black, citecolor=black, urlcolor=black]{hyperref} 
\usepackage{array}
\usepackage{orcidlink}
\usepackage[caption=false,font=footnotesize,labelfont=rm,textfont=rm]{subfig}
\usepackage{textcomp}
\usepackage{stfloats}
\usepackage{url}
\usepackage{verbatim}
\usepackage{graphicx}
\usepackage{cite}
\usepackage{gensymb}
\usepackage{multirow}
\usepackage{booktabs}
\usepackage{colortbl}
\usepackage{makecell}
\usepackage{enumitem}
\usepackage{xcolor}
\usepackage{tikz}
\usepackage{amsmath}
\usepackage[linesnumbered,ruled,vlined,algo2e]{algorithm2e}
\usepackage[ruled, vlined, linesnumbered]{algorithm2e}
\usepackage{algorithm}
\usepackage{algpseudocode}
\usepackage{multirow}
\usepackage{color}
\usepackage{filecontents}
\usepackage{subfig,graphicx}
\usepackage{xspace}

\newcommand\SystemName{\textsc{MetaAttack}\xspace}

\newcommand\cparagraph[1]{\vspace{1.2mm}\noindent \textbf{#1.}}
\begin{document}
% \pagewiselinenumbers% 按页重新编号 
% \switchlinenumbers

% \linenumbers
\title{A Portable and Stealthy Inaudible Voice Attack Based on Acoustic Metamaterials}
% Zhiyuan Ning$^{*}$, Juan He$^{*}$, Zhanyong Tang, Weihang Hu, Xiaojiang Chen

\author{Zhiyuan Ning$^{*\orcidlink{0009-0009-4416-2080}}$,
        % <-this % stops a space
Juan He$^{*\orcidlink{0000-0003-1856-6347}}$,
        % <-this % stops a space
Zhanyong Tang$^{\orcidlink{0000-0002-4333-2334}}$, 
        % <-this % stops a space
Weihang Hu$^{\orcidlink{0009-0009-5841-2079}}$, 
        % <-this % stops a space
and Xiaojiang Chen$^{\orcidlink{0000-0002-1180-6806}}$

% \thanks{}
\thanks{This work was supported in part by the National Natural Science Foundation of China (NSFC) under Grant 62372373 and in part by the Shaanxi Province “Engineers + Scientists” Team Building Program under Grant 2023KXJ-055. (\textit{* Zhiyuan Ning and Juan He contributed equally to this work. Zhanyong Tang is the corresponding author.})

Zhiyuan Ning is with the School of Information Science and Technology, Northwest University, Xi’an 710127,  China (e-mail: ningzhiyuan@stumail.nwu.edu.cn). 

Juan He is with the School of Information Science and Technology, Northwest University, Xi’an 710127,  China (e-mail: hejuan@stumail.nwu.edu.cn).

Zhanyong Tang is with the Shaanxi Key Laboratory of Passive Internet of Things and Neural Computing, with the Xi’an Key Laboratory of Advanced Computing and Software Security, and with the School of Information Science and Technology, Northwest University, Xi’an 710127, China  (e-mail: zytang@nwu.edu.cn).

Weihang Hu is with the School of Information Science and Technology, Northwest University, Xi’an 710127,  China (e-mail: huweihang\_nwu@163.com).

Xiaojiang Chen is with the School of Information Science and Technology, Northwest University, Xi’an 710127,  China (e-mail: xjchen@nwu.edu.cn).

This paper has supplementary downloadable material available at http://ieeexplore.ieee.org, provided by the author. The material includes two video demonstrations. Contact zytang@nwu.edu.cn for further questions about this work.
}
}

\markboth{Journal of \LaTeX\ Class Files,~Vol.~18, No.~9, September~2020}%
{How to Use the IEEEtran \LaTeX \ Templates}
 \IEEEpubid{0000--0000/00\$00.00~\copyright~2021 IEEE}
\maketitle

\begin{abstract}
We present \SystemName, the first approach to leverage acoustic metamaterials for inaudible attacks for voice control systems. Compared to the state-of-the-art inaudible attacks requiring complex and large speaker setups, \SystemName achieves a longer attacking range and higher accuracy using a compact, portable device small enough to be put into a carry bag. These improvements in portability and stealth have led to the practical applicability of inaudible attacks and their adaptation to a wider range of scenarios. We demonstrate how the recent advancement in metamaterials can be utilized to design a voice attack system with carefully selected implementation parameters and commercial off-the-shelf components. We showcase that \SystemName can be used to launch inaudible attacks for representative voice-controlled personal assistants, including Siri, Alexa, Google Assistant, XiaoAI, and Xiaoyi. The average success rate of all assistants is 76\%, with a range of 8.85 m.
\end{abstract}


\section{INTRODUCTION}

\ifx\allfiles\undefined	

\else
\fi

\IEEEPARstart{V}{oice} control is a natural way for human-computer interactions, commonly found in mobile and smart home applications~\cite{voicecloak,hu2022accear,hu2022milliear,hu2022towards,hu2023mmecho,zhang2024echolight,yang2024rf,wang2024wireless,handling,watching}.
Given its capacity to control a broad spectrum of tasks -- from making phone calls to managing devices like surveillance cameras and even initializing bank transfers -- voice control systems have increasingly become the target of  attacks~\cite{r7,dol,r9,r100,r101,Fencesitter,Magbackdoor,voicecloak}.

Recent research has demonstrated that the voice control mechanisms in mainstream personal assistant systems, like Amazon's Alexa and Apple's Siri, are susceptible to inaudible attacks~\cite{r7,dol,r9}. Such attacks modulate voice commands onto ultrasonic carriers, generating commands imperceptible to human ears but can be accepted by voice control systems. Despite some levels of success in specific scenarios, existing portable inaudible attacking methods (such as relying on lithium batteries and using small speaker arrays to transmit attack signals)~\cite{r9,perturbation,dol} have a short attack range of up to 3 m. However, a close-in attack within 3~m can be easily detected by many vigilant individuals~\cite{lipread}. Extending this range by simply increasing transmission power would lead to audible noise, thereby compromising the stealthiness of the attack. To address this limitation, more recent work, as represented by LipRead~\cite{lipread}, divides the spectrum of inaudible commands to reduce audible leakage and employs a large array of 61 speakers to extend the attack distance up to 7.62 meters. DolphinAttack~\cite{longdol}, targeting the iPhone X, uses 40 speakers along with external power supplies and amplifiers to achieve a range of 19.8 meters. In comparison, although these methods can significantly expand the attack distance, their reliance on bulky equipment and external power sources greatly reduces portability, limiting their applicability in many real-world public settings.

\begin{figure}[t!]
    \centering
    \includegraphics[width=0.95\linewidth]{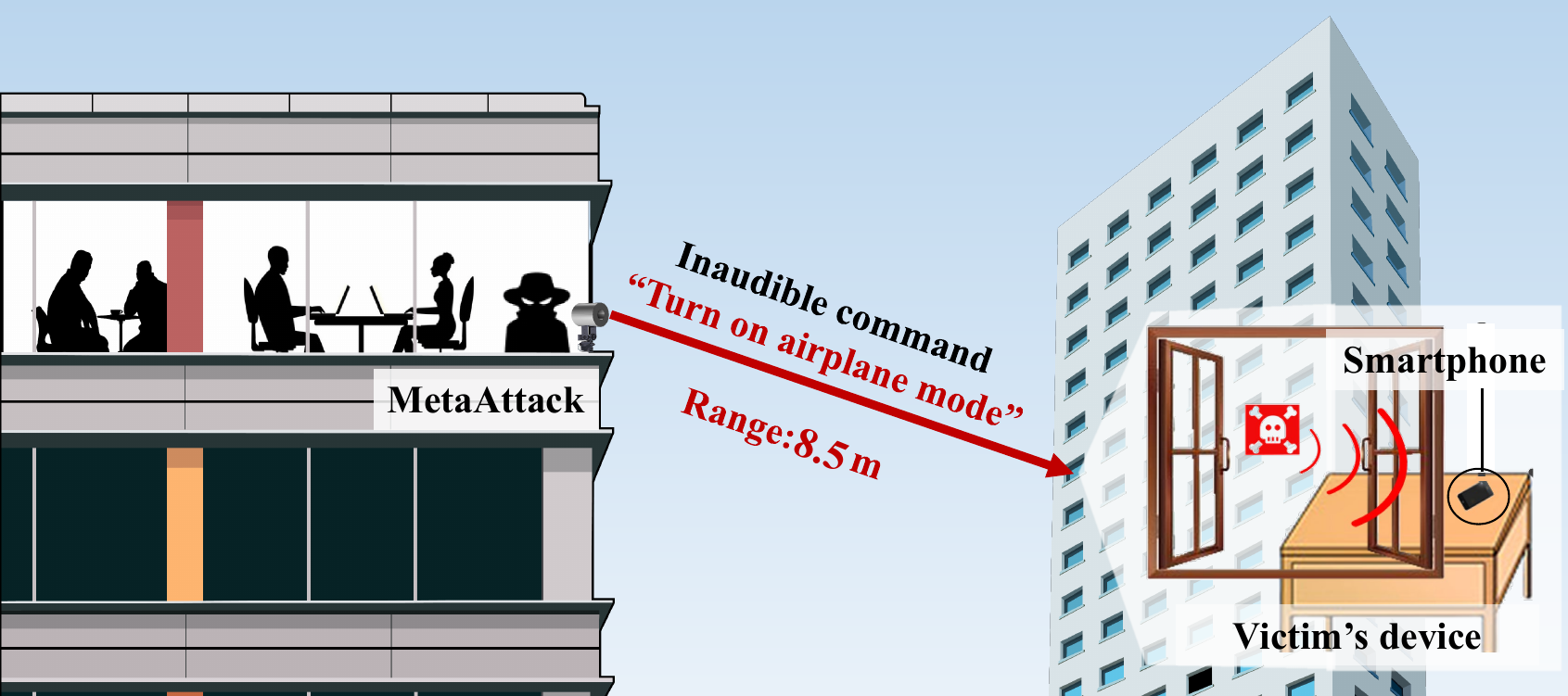}
    \caption{A possible attacking scenario of \SystemName where an attacker can launch an inaudible attack in a public place to remotely control the target device in another building. In this scenario, the attack device must be stealthy and does not require the use of other continuously powered devices such as DC power supply.}
    \label{A possible attacking scenario}
    \vspace{-10pt}
\end{figure}

This paper demonstrates the feasibility of executing long-range, inaudible attacks through a compact audio modulation system that can easily fit into a modest-sized travel bag. We introduce \SystemName, the first inaudible attack that leverages the emergent acoustic metamaterials technology~\cite{r75,r76,r77,r78,r79,r106}. Acoustic metamaterials are engineered to enable the manipulation, control, and transformation of acoustic waves in ways impossible to conventional materials. These metamaterials' dimensions are smaller than the wavelengths of the sound waves they interact with. By carefully designing the elements' geometry, size, shape, and spatial arrangement, \SystemName can precisely control the acoustic wave propagation via compact devices~\cite{r11,r12,r13}. For example, Fig. \ref{A possible attacking scenario} depicts a possible attack scenario where \SystemName is covertly placed in a public place next to a window to target a device located in another building within 8.5~m.
\IEEEpubidadjcol

While acoustic metamaterials are shown to be useful in other domains~\cite{r14,r15,r16,r17}, there is no prior work for using them to build a practical inaudible attack system. In our work, we introduced two changes to improve the metamaterial's functionality. First, we refined the internal structure to expand its filtering capabilities, thus minimizing audible leakage during inaudible attacks. Second, we converted the metamaterial into a piston sound source ~\cite{r86,r118} by carefully designing the incident sound wave's shape, increasing the energy intensity of inaudible commands, and extending the attack range. We call our enhanced design a multi-functional acoustic metamaterial (MAM), as it not only filters out audible leakage but also \emph{amplifies the energy of inaudible commands}. With these advancements and a compact design, \SystemName can execute long-range inaudible attacks while remaining highly concealed.

We showcased that \SystemName can be implemented with low-cost, commercial, off-the-shelf components like Raspberry Pi, amplifiers, loudspeakers, servo, and a 3D-printed MAM. We thoroughly evaluate \SystemName under different practical scenarios under different attack ranges, tasks, noise levels, obstacles, and targeting different target devices. We compare \SystemName against three prior inaudible attacks~\cite{dol,r9,lipread}. Experimental results show that \SystemName is robust under different evaluation scenarios and tasks and can successfully launch an inaudible attack beyond 8 m range with a noise level of up to 65 dB. Compared with prior attacking methods, \SystemName achieves a 76\% average success rate under a 8.85 m attack range, making it the smallest device to achieve such performance.

This paper makes the following contributions:
\begin{itemize}[leftmargin=*]
\item We present the first acoustic metamaterial-based inaudible attacks and released a complete process demonstration of \SystemName \cite{MetaAttack,OutdoorMetaAttack} and open-sourced the code \cite{feedback};
\item We demonstrate how acoustic metamaterial can be repurposed to improve the capability of inaudible attacks;
\item We showcase how a practical attacking device can be built from commercial off-the-shelf components and discuss the potential countermeasures.

\end{itemize}

\ifx\allfiles\undefined

\else
\fi
\section{BACKGROUND} \label{chap:2}

\subsection{Inaudible Attacks}
Inaudible attacks leverage ultrasonic frequencies (sounds above 20 kHz), which the microphones in smart devices can often detect even though humans cannot hear them. The concept behind these attacks is to issue commands to voice-activated devices without the device owner's knowledge. 

In an inaudible attack, the microphone receives input signals that include not only the audible-frequency voice control signal but also a high-frequency carrier signal. Due to the microphone's non-linear response, these input signals generate additional non-linear components in the output, providing the conditions for recognizing and transmitting the attack signal. Eq.\ref{eqn-1} represents the composition of the input signal ${S_{in}}(t)$, which is as follows:
\begin{equation}\label{eqn-1} 
{S_{in}}(t) = {V_{attack}}\left( t \right)\cos \left( {2\pi {f_{carrier}}t} \right) + \cos \left( {2\pi {f_{carrier}}t} \right)
\end{equation}
where ${V_{attack}}\left( t \right)$ is the voice control signal, typically a low-frequency control signal that the device can recognize. $\cos \left( {2\pi {f_{carrier}}t} \right)$ is the high-frequency carrier signal, with a frequency of ${f_{carrier}}$, which is beyond the human hearing range (typically above 20 kHz). The role of the carrier signal is to add additional frequency components to the attack signal, utilizing the non-linear characteristics of the speaker array when outputting the signal to generate additional components. The attack signal output by the speaker array, ${S_{attack}}(t)$, is represented by Eq.\ref{eqn-2}:
\begin{equation}\label{eqn-2} 
\begin{aligned}
S_{attack}(t) & = S_{h}+\frac{A_{2}}{2}\left(1+V_{attack}^{2}(t)+2 V_{attack}(t)\right)
\end{aligned}
\end{equation}
where $S_{h}$ represents background noise or other unrelated signals that are not associated with the attack. The term $\frac{A_{2}}{2} \left(1 + V_{attack}^{2}(t) + 2 V_{attack}(t) \right)$ represents the additional signal components introduced by non-linear effects. In this case, ${V_{attack}}\left( t \right)$ can be recognized by the target's microphone, enabling the successful execution of the inaudible attack.

By utilizing the modulation method described above, an adversary encodes commands in ultrasonic frequencies that the microphones of nearby voice-activated devices can pick up. For example, these commands could be ``turn off the surveillance camera'' if the system is connected to smart home security, ``make a purchase'' and ``transfer money'' if linked to financial services, ``sending counterfeit messages'', or ``calling number X'' for eavesdropping during a meeting. The ultrasonic commands are transmitted over the air, potentially from a loudspeaker system or a specialized device, and received by the target device. Because these sounds are inaudible to humans, the device owner remains unaware that their device is being controlled remotely. However, the voice recognition of the targeted device can process these sounds as normal voice commands and execute the action without the user’s knowledge or consent, posing a significant vulnerability for voice-controlled smart devices.

\subsection{Threat Model}\label{Threat Model2.2} 
\cparagraph{Portable, miniaturized and stealthy inaudible attack} We focus on implementing an inaudible acoustic attack that is highly portable, miniaturized, and can be executed outside of detection range. Miniaturization and portability mean that the system does not rely on signal amplifiers powered by DC power supplies or bulky speaker arrays, and can be easily carried in a handbag or small backpack, allowing the attacker to operate discreetly in public spaces. The so-called detection range refers to a range in which the target is unlikely to detect the presence of the attacker. According to related studies \cite{r99,privacy2,privacy3}, a distance greater than 3 meters is considered a \textit{safe distance} when interacting with strangers, typically not raising suspicion.

\cparagraph{No victim device interaction during attack} We assume that during the attack, the victim's device is in close proximity but not actively in use, with its voice assistant idle and the screen not being viewed by the victim. This assumption aligns with those made in prior audio-based attack studies, such as Dolphin~\cite{dol,longdol}, NUIT~\cite{nuit} and GhostTalk~\cite{r101}, and reflects realistic scenarios, such as when the victim places the device aside or face-down on a desk to focus on tasks like studying or attending a meeting.

\cparagraph{Obtain the victim's voiceprint} We assume the attacker can synthesize the target's voice using, e.g., deep-learning-based voice generation from a few training samples~\cite{dol}. We believe this assumption is reasonable in many day-to-day scenarios where the target's voice samples can be collected from public places, a multi-user meeting, or audio and video clips available to the attacker.

\cparagraph{Attack victim device in pocket} We consider the scenario where the victim may place their device in their pocket for convenience while out, during which the pocket could physically interfere with the transmission of attack commands. For this scenario, we have verified the system's effectiveness in Section~\ref{S-2}.

\subsection{Practical Limitations of Prior Methods}

Launching inaudible voice commands is achieved by modulating the control signal, but high-power operation of speakers can cause nonlinear effects, leading to audible leakage, which limits the attack range. Previous studies \cite{dol,r9} avoided leakage by limiting the device power, but this restricted the attack range to within 3 meters. To achieve longer-range attacks, LipRead \cite{lipread} divides the frequency spectrum into low-frequency parts to reduce human perception, but it requires 61 speakers, increasing the risk of detection, and spectrum segmentation may reduce the accuracy of commands.
Similarly, Yan et al. \cite{longdol} extended the attack range to 19.8 meters, but this requires a signal generator and DC power supply, which does not meet the needs of portable attacks, and requires 40 speakers.
To clearly illustrate the limitations of existing research, we provide a comparison in Table \ref{Compare}. Most studies focus on voiceprint recognition attacks \cite{lipread,r7,r9,dol}, but these attacks often require close proximity or a large number of auxiliary devices, lacking stealth. Therefore, a successful attack requires a balance between portability and attack range.

\begin{table*}
    \scriptsize
    \caption{Comparison of prior inaudible attack research in terms of attack range, portability and stealthy}
    \label{Compare}
    % \vspace{-3mm}
    \renewcommand{\arraystretch}{1}
    \centering
    \begin{tabular} { m{1.4cm}  c m{3.2cm}  m{3.2cm}  m{0.8cm} m{1.7cm} m{0.8cm} m{0.8cm}}
    \toprule
    \textbf{\makecell[c]{System}} & \textbf{\makecell[c]{Setup}} & \textbf{\makecell[c]{Power supply}}& \textbf{\makecell[c]{Signal generator}}&
    \textbf{\makecell[c]{Speakers}}&
    \textbf{\makecell[c]{Attack range}}&
    \textbf{\makecell[c]{Portable}}&
    \textbf{\makecell[c]{Stealthy}}\\

    \midrule
        \makecell[l]{DolphinAttack\\(Long-range\\ System) \cite{longdol}} & \makecell[c]{\begin{minipage}[b]{0.27\columnwidth}
		\raisebox{-.3\height}{\includegraphics[width=\linewidth]{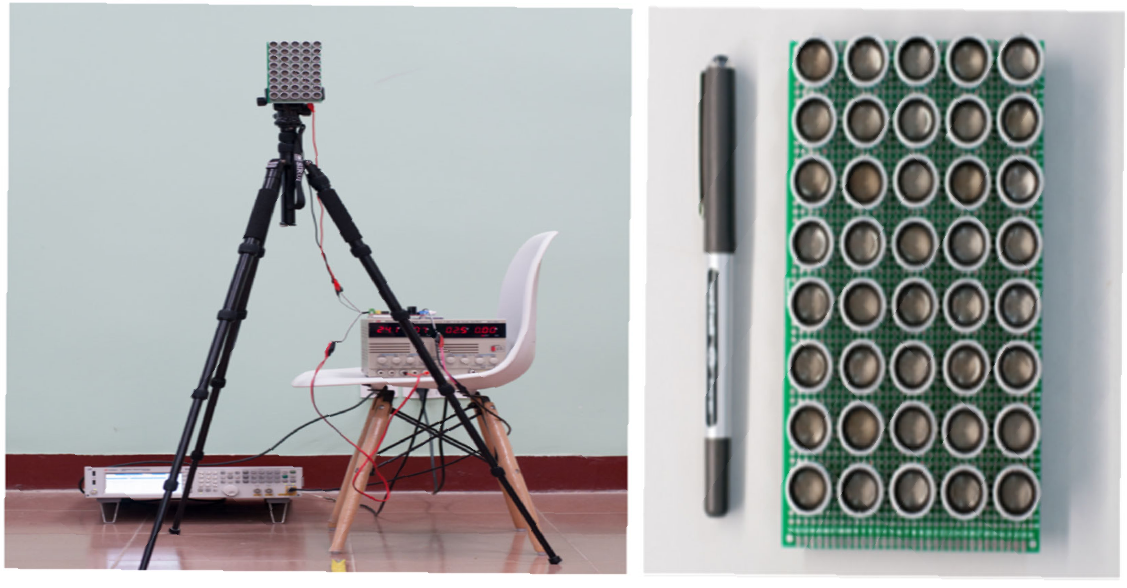}} \end{minipage}} 
        &  \makecell[c]{Yes \\($Width*Hight*Depth \approx$ \\ $26*16.5*30~cm \approx$ \\ $12~MacBook~Pro)$} 
        & \makecell[c]{Yes \\($Width*Hight*Depth \approx$ \\ $42.6*8.8*48.3~cm \approx$ \\ $16~MacBook~Pro)$} 
        &  \makecell[c]{40}
        &\makecell[c]{19.8m (iPhoneX)\\10m (iPhoneSE)}
        & \makecell[c]{No}
        & \makecell[c]{No}
        \\
 
\hline       
        LipRead \cite{lipread} & \makecell[c]{ \begin{minipage}[b]{0.27\columnwidth}
		\raisebox{-.3\height}{\includegraphics[width=\linewidth]{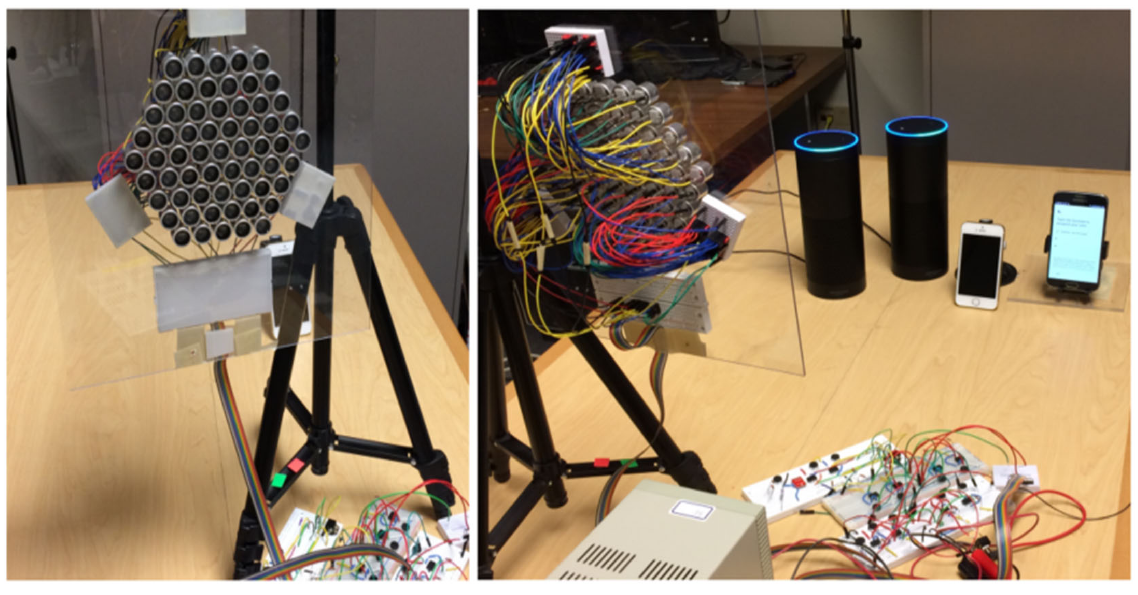}} \end{minipage}}  
        &  \makecell[c]{Yes \\($Width*Hight*Depth \approx$ \\ $21.3*31.2*36~cm \approx$ \\ $22~MacBook~Pro)$} 
        & \makecell[c]{Yes \\($Width*Hight*Depth \approx$ \\ $26.1*10.3*30.3~cm \approx$ \\ $7~MacBook~Pro)$} 
        &  \makecell[c]{61}
        & \makecell[c]{7.62m}
        & \makecell[c]{No}
        & \makecell[c]{No}
        \\ 

\hline

        BackDoor \cite{r7} & \makecell[c]{\begin{minipage}[b]{0.27\columnwidth}
		\raisebox{-.5\height}{\includegraphics[width=\linewidth]{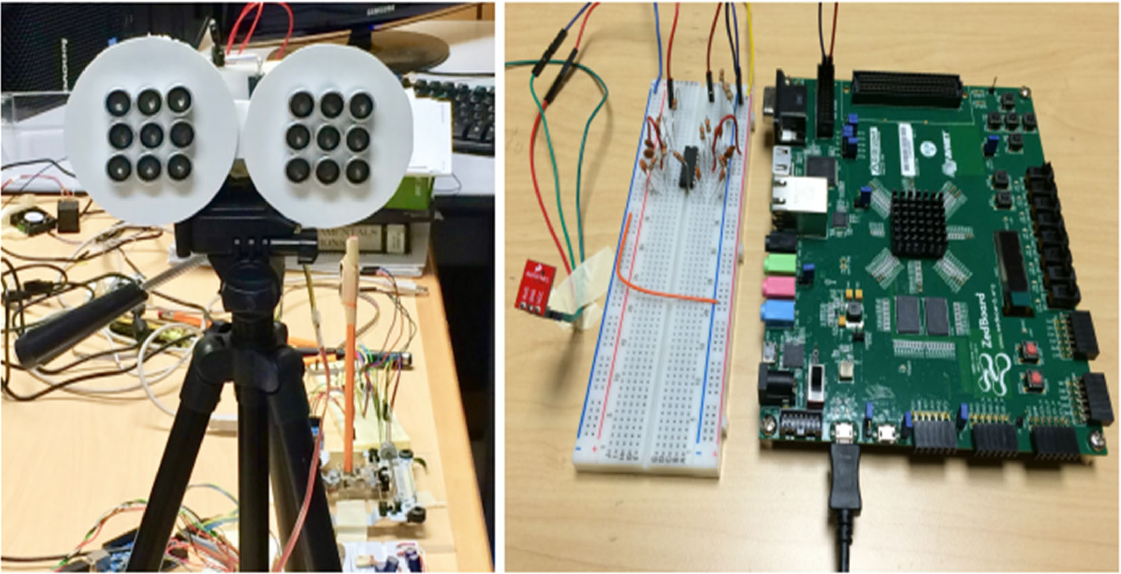}} \end{minipage}} 
        &  \makecell[c]{Yes \\($Width*Hight*Depth \approx$ \\ $21.3*31.2*36~cm \approx$ \\ $22~MacBook~Pro)$} 
        & \makecell[c]{Yes \\($Width*Hight*Depth \approx$ \\ $26.1*10.3*30.3~cm \approx$ \\ $7~MacBook~Pro)$} 
        &  \makecell[c]{18}
        &\makecell[c]{3.5m}
        & \makecell[c]{No}
        & \makecell[c]{No}
        \\

\hline
        EchoAttack \cite{r9} & \makecell[c]{\begin{minipage}[b]{0.27\columnwidth}
		\raisebox{-.5\height}{\includegraphics[width=\linewidth]{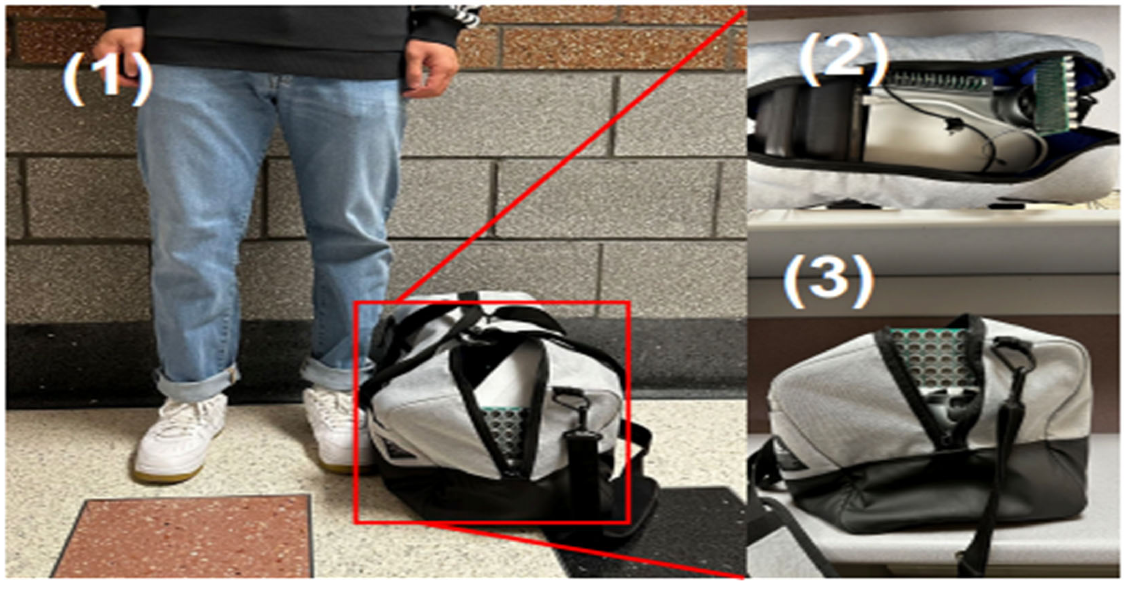}} \end{minipage}} &  \makecell[c]{No} 
        & \makecell[c]{Yes \\($Width*Hight*Depth \approx$ \\ $26.1*10.3*30.3cm \approx$ \\ $7~MacBook~Pro)$}
        &  \makecell[c]{40}
        &\makecell[c]{2.23m}
        & \makecell[c]{Yes}
        & \makecell[c]{Yes}
        \\

\hline
        VRIFLE \cite{perturbation} & \makecell[c]{\begin{minipage}[b]{0.27\columnwidth}
		\raisebox{-.5\height}{\includegraphics[width=\linewidth]{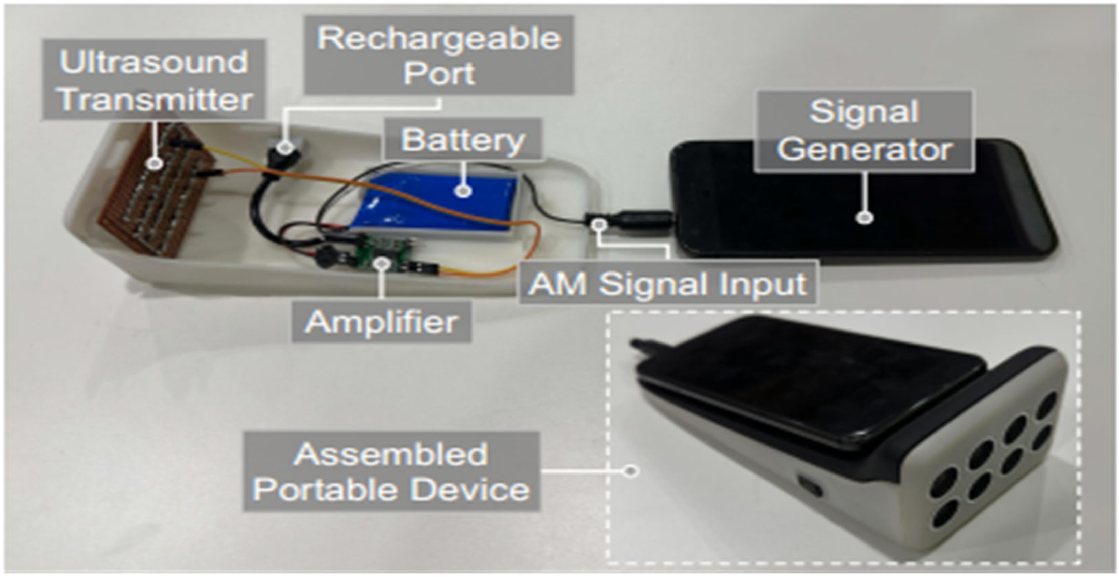}} \end{minipage}}
        &  \makecell[c]{No}
        & \makecell[c]{No}
        &  \makecell[c]{8}
        &\makecell[c]{1.8m}
        & \makecell[c]{Yes}
        & \makecell[c]{Yes}
        \\
 
\hline       
        \makecell[l]{DolphinAttack\\(Portable\\ System) \cite{dol}} & \makecell[c]{\begin{minipage}[b]{0.27\columnwidth}
		\raisebox{-.5\height}{\includegraphics[width=\linewidth]{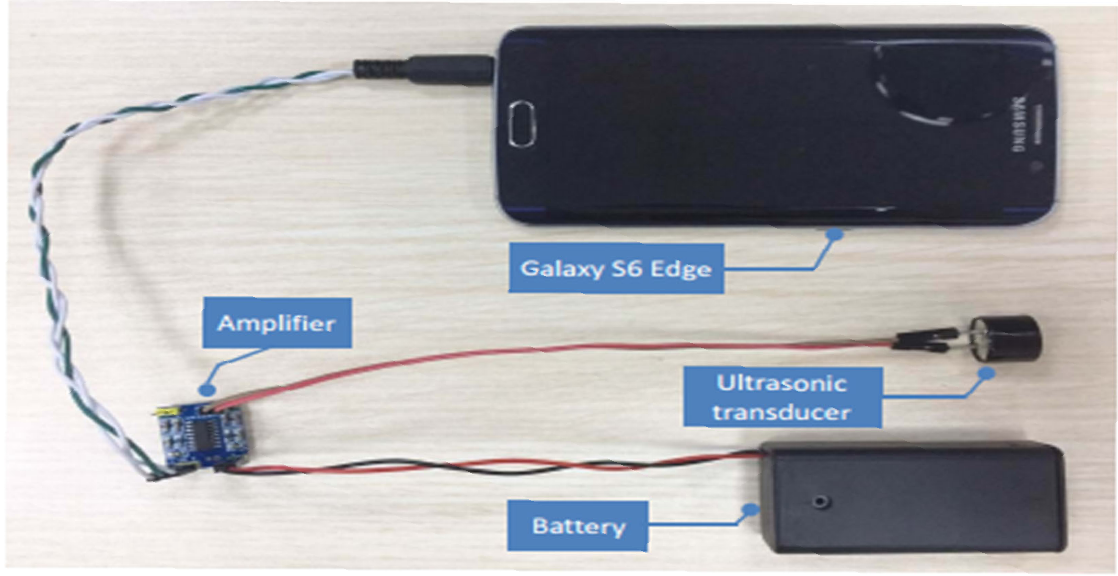}} \end{minipage}}
        &  \makecell[c]{No}
        & \makecell[c]{No}
        &  \makecell[c]{1}
        &\makecell[c]{0.27m}
        & \makecell[c]{Yes}
        & \makecell[c]{Yes}
        \\

    \bottomrule
    \end{tabular}
    \vspace{-4mm}
\end{table*}

\subsection{Motivations}
\begin{figure}
    \centering    \includegraphics[width=0.9\linewidth]{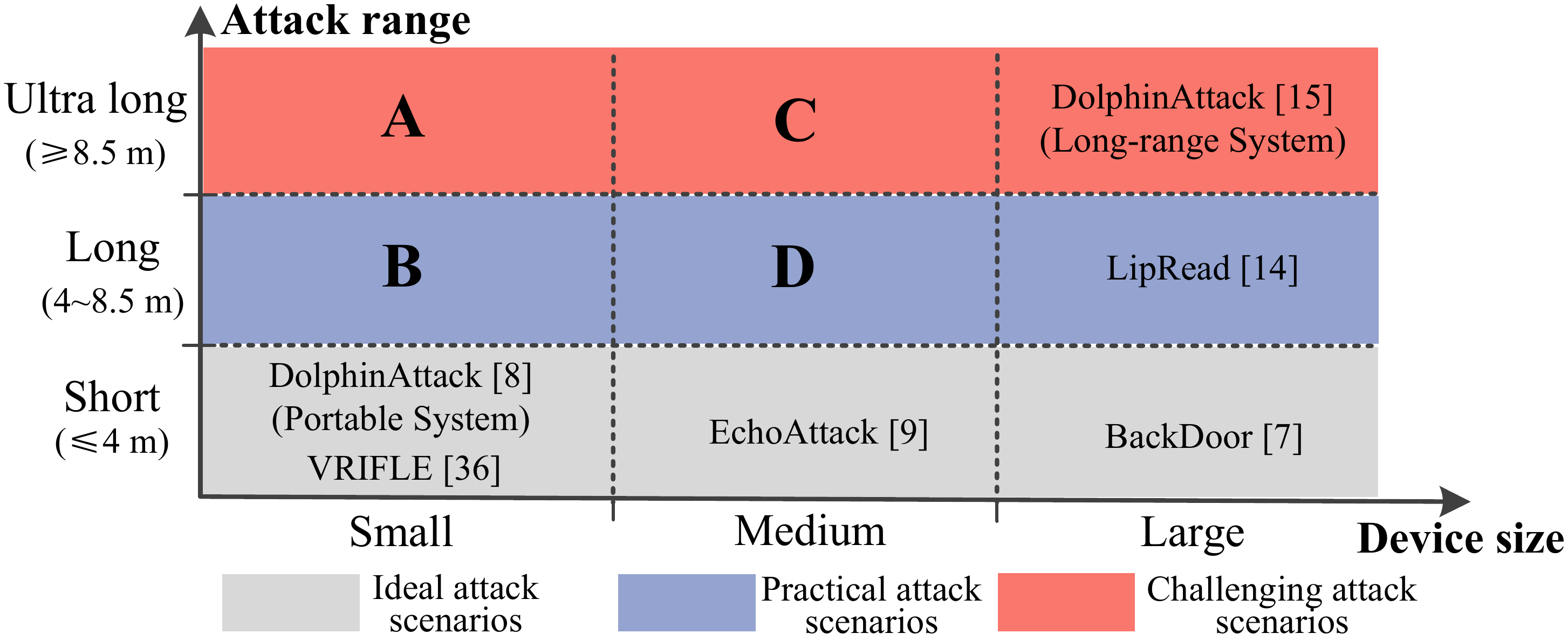}
    \caption{Attack Range/Device Size quadrant chart. (Large: with DC power supply, signal generator and large speaker array. Medium: with a signal generator and large speaker array. Small: without above devices.)}
    \label{Inaudible research}

\end{figure}

Most of the attack scenarios presented in previous studies are based on the assumption that the device to be attacked is stationary in a specific location \cite{lipread,r7,longdol,dol}, or that a large speaker array is used if the distance is too great \cite{lipread,longdol} - assumptions that are not always true. This is because in many public environments, using external power through amplification devices could easily expose the malicious intent behind an attack, often leading to early detection before it is initiated. As shown in Table \ref{Compare}, if the attacking system lacks DC power support, its attack range is limited\cite{lipread,longdol}. In this paper, for parts A and B of Fig~\ref{Inaudible research}, we aim to design an attack device that is portable and concealable, without the need for an external DC auxiliary power supply, and that can be implemented over long distances.

\subsection{Metamaterials}
We present the first inaudible attack using acoustic metamaterials. Our key insight is that a carefully engineered metamaterial can be designed with structures that interact with sound waves in specific ways, allowing them to block or absorb sound at certain frequencies while letting other frequencies pass through \cite{r112,r113}. This means we can employ metamaterials to create filters that target specific sound frequencies while allowing airflow, enhancing the power of inaudible commands and minimizing audible leakage. In this work, we leverage metamaterials to design a compact, passive device, offering two main benefits over existing inaudible attacking methods. First, the MAM preserves the accuracy of inaudible commands by filtering audible sounds without splitting their spectrum. Secondly, it enables using a few loudspeakers to replicate the effect of a larger array, facilitating covert and long-range attacks.

\section{OVERVIEW of OUR APPROACH} \label{chap:5}

\begin{figure}
    \centering    \includegraphics[width=0.9\linewidth]{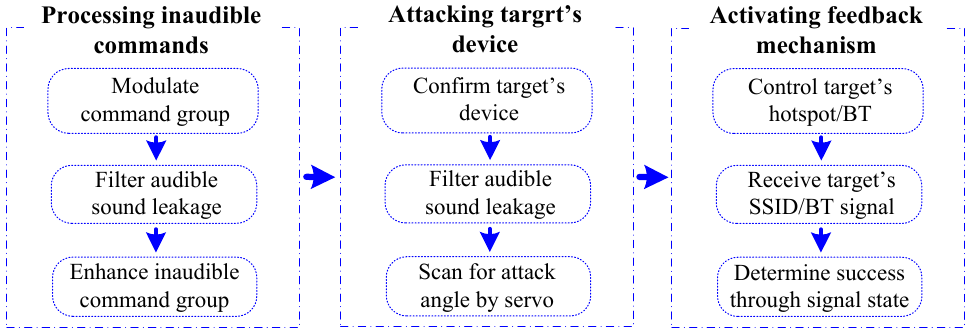}
    \caption{Overview of \SystemName with three steps.}
    \label{MetaAttack System}
    \vspace{-10pt}
\end{figure}

As shown in Fig. \ref{MetaAttack System}, \SystemName is a 3-step inaudible attack, described in detail below. 

\cparagraph{Processing inaudible commands}
The initial phase of our attack involves processing and modulating voice commands. We first capture the victim’s voice and synthesize their voiceprint (see Section \ref{Threat Model2.2}). Then, we modulate the voice commands generated by Google Text-to-Speech AI \cite{Google} together with the synthesized voiceprint onto ultrasonic carrier waves using the method described in Section \ref{chap:2}, generating inaudible commands \cite{lipread}. Each command group includes four types of inaudible instructions: \textit{Mute commands} (e.g., “speak 6\%” for Siri or “turn the volume to 1” for other devices \cite{nuit,surfingattack}, denoted as $MuteCmd$), \textit{Attack commands} ($AttackCmd$), \textit{Feedback commands} (e.g., “turn on hotspot” and “turn off hotspot,” denoted as $FeedbackCmd_1$ and $FeedbackCmd_2$) and \textit{Reset commands} (e.g., “speak 50\%” for Siri or “turn the volume to 5” for other devices, denoted as $ResetCmd$).

\cparagraph{Attacking target device} To accurately target the device, \SystemName uses a servo motor to rotate in 12° steps within a specified range, gradually aligning with the victim. The system first sends the $MuteCmd$ to mute the device \cite{nuit,surfingattack}, then sends the $AttackCmd$ to carry out the attack, activates the feedback mechanism to confirm the success of the attack, and finally uses the $ResetCmd$ to restore the device's volume. %Additionally, leveraging recent research, the system can implement covert deletion of call or text message records \cite{delete1,delete2,delete3}, further enhancing the stealth of the attack.
Leveraging recent research findings\cite{ delete1,delete2,delete3}, our system has successfully implemented the covert deletion of call and text message records, thereby significantly enhancing the attack's stealth capabilities.

To ensure the reliability of the attack execution, each command is sent consecutively five times. Since the execution time of each command is only a few seconds, repeated transmissions do not introduce any significant negative effects. Leveraging MetaAttack's high single-command success rate of 76\% at a distance of 8.85 meters, this strategy raises the theoretical final success rate of each command to 99.76\%. In addition, we conducted 50 rounds of complete feedback processes to validate the execution success rate of each attack stage. Among them, the $MuteCmd$, $AttackCmd$, $FeedbackCmds$ and $ResetCmd$ each achieved a 98\% success rate, with only one failure per command. The overall success rate of the full attack chain was 94\%, with only three failures, demonstrating the high efficiency and stability of the attack process.

\cparagraph{Activating feedback mechanism}
This feedback mechanism uses a \textit{Raspberry Pi} to monitor the state changes of the victim device after receiving a wireless switching command to determine whether the attack was successful. If the victim device’s hotspot signal undergoes a state transition of \texttt{appearance}–\texttt{disappearance}–\texttt{reappearance}, the command execution is considered successful. 

The reason for using the sequence \texttt{appearance} – \texttt{disappearance} – \texttt{reappearance} is that relying solely on \texttt{appearance} – \texttt{disappearance} may lead to false positives. If a non-victim device happens to activate its hotspot between the two feedback commands, the system may incorrectly interpret this as an attack success and prematurely terminate the attack process. To address this, we introduce \texttt{appearance} – \texttt{disappearance} – \texttt{reappearance} to confirm the continuity of the response, as it is highly unlikely for a non-victim device to switch its hotspot twice within a short period. To further reduce false positives, we incorporate \textit{iwlist} \cite{iwlist} to monitor hotspot SSIDs and signal strength in real time, addressing the random appearance or disappearance of weak signals from non-victim devices. Typically, such weak signals show gradual changes in strength when appearing or disappearing, whereas the signal from the victim device triggered by $FeedbackCmd$ exhibits sudden changes. Based on this difference, we filter out SSIDs with abrupt signal strength transitions, effectively eliminating interference and improving detection accuracy.

\begin{figure}[t]
    \centering
    \includegraphics[width=1\linewidth]{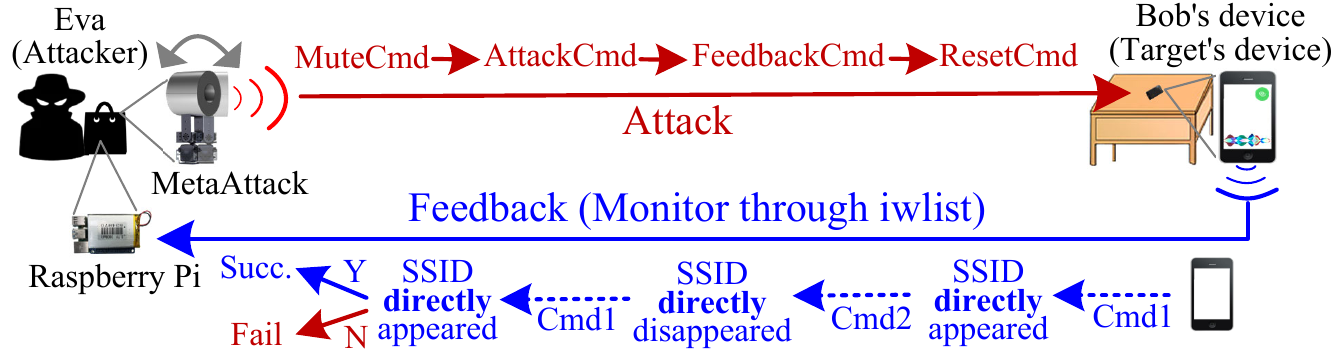}

    \caption{\SystemName Overview: $MuteCmd$ and $AttackCmd$ initiate the attack, $FeedbackCmd$ confirms success and $ResetCmd$ restores the target.}
    \label{Feedback overview}
\end{figure}

As shown in Algorithm \ref{Algorithm} and Fig. \ref{Feedback overview}, to illustrate the feedback mechanism for smartphones, we break it down into four sequential steps:

\cparagraph{Step 1} After transmitting the $MuteCmd$ and $AttackCmd$ (line~\ref{alg:3}), the Raspberry Pi uses iwlist to record the identified hotspot devices and stores them in the \textit{L1} list (line~\ref{alg:13}). Then, it sends the $FeedbackCmd$ and stores the new records in the \textit{L2} list (line~\ref{alg:14}). To prevent the system from repeatedly executing $MuteCmd$ and $AttackCmd$ while waiting for the success of $FeedbackCmd$, we introduced the concept of a time window. Considering that the system needs a certain amount of time to receive the victim device's hotspot signal via the Raspberry Pi, no other commands will be sent within a time window after sending $FeedbackCmd$, thereby avoiding the repeated execution of commands.)

\cparagraph{Step 2} The system continues to send $FeedbackCmd_2$ and stores the received hotspot information in list \textit{L3} (line \ref{alg:17}). If there are discrepancies in the hotspot signals ($dif_1$) between the \textit{L2} and \textit{L3} lists, the process moves to \textbf{Step 3} (line \ref{alg:22}). Otherwise, the \SystemName is rotated to the next angle (line \ref{alg:23}).

\cparagraph{Step 3} \SystemName will sequentially send $FeedbackCmd_1$ and then instruct the Raspberry Pi to retrieve the \textit{L4} hotspot list~(line ~\ref{alg:26}). If there are differences in the hotspot signals ($dif_2$) between the \textit{L3} and \textit{L4} lists, a successful stealth attack is detected~(lines \ref{alg:27} - \ref{alg:31}). 
If not, the \SystemName will be adjusted to the next angle.

\cparagraph{Step 4} To maintain the stealth of the attack, this step focuses on restoring the victim device to its original state before the attack~(line \ref{alg:37}).

\cparagraph{Optimization of feedback strategies in different scenarios} Considering the time overhead of the feedback algorithm and the limited angular coverage of a single attack, we have designed additional strategies to improve attack efficiency: for attacks on visible targets, the attacker can manually align the system with the target, completing the attack within 1 minute.

In addition, for commands that provide direct feedback to the attacker (such as making a call for eavesdropping), we immediately confirm whether the call was successfully connected after the command is executed. If the connection is successful, the attack is stopped via SSH connection to the Raspberry Pi; if it fails, the subsequent feedback is halted and the system moves to the next angle to reduce time overhead. By tailoring strategies to different environments, the system achieves efficient and flexible attack execution.

To confirm the feasibility of our attack process, we have released a complete process demonstration \cite{MetaAttack} and open-sourced the feedback code \cite{feedback}.
            
\begin{algorithm2e}[!t]
% \scriptsize
\footnotesize
    \SetAlgoLined %显示end
	\caption{Attack overview}%算法名字
        \label{Algorithm} %用于文内引用的标签
        \SetKwFunction{FB}{feedbackMechanism}
        \SetKwFunction{one}{Step1}
        \SetKwFunction{two}{Step2}
        \SetKwFunction{three}{Step3}
        \SetKwFunction{four}{Step4}
        \SetKwProg{Fn}{Function}{:}{}
 
            \ForEach{$angle$}{ \label{alg:2}

            $Send\ MuteCmd,\ AttackCmd,\ activate\ iwlist$  \\  \label{alg:3}

             $\FB{}$\\
            \If{$ \FB{}\ \textbf{==}\ true\ $ }{
            \textbf{break} \\}
            } \label{alg2:7}

         \label{alg2:8}
        
        \Fn{\FB{}}{

            $\one{}$\\

           $\textbf{return}\ \two{$L_1, L_2$} $\\
            }

            \Fn{\one{}}{
            \tcp{$Attempt\ to\ turn\ on\ victim's\ hotspot$}
            $Receive\ hotspot\ list\ L_1$ \label{alg:13}\\
            $Send\ FeedbackCmd_1$\\
            $Wait\ time\ window\ and\ receive\ hotspot\ list\ L_2 $ \label{alg:14}\\ 
            $\textbf{return}\ L_1, L_2 $
            }
            \Fn{\two{$L_1, L_2$}}{
            \tcp{$If\ any\ hotspot\ signal\ disappears,\ moves\ to\ Step3$}
            $Send\ FeedbackCmd_2,\ receive\ hotspot\ list\ L_3 $ \label{alg:17}\\
            $dif_1 = L_2-L_3$ \label{alg:18}\\
            \eIf{$L_2 \neq L_3$}{\label{alg:19}
            % $\three{$L_1,L_3,dif_1$}$ \label{alg:20}\\
            $\textbf{return}\ \three{$L_1,L_3,dif_1$} $\label{alg:22} }{
            $\textbf{return}\ false$ \label{alg:23}
            }
            }
            
            \Fn{\three{$L_1,L_3,dif_1$}}{ \label{alg:33}
            \tcp{$If\ hotspot\ signal\ reappears,\ attack\ is\ successful$}
            $Send\ FeedbackCmd_1$\\
            $Wait\ time\ window\ and\ receive\ hotspot\ list\ L_4 $ \label{alg:26}\\
            $dif_2 = L_4-L_3$ \label{alg:27}\\
            $Target\ list = dif_1 \cap dif_2$ \label{alg:28}\\
            
            \four{$Target\ list, L_4, L_1$}\\
            \textbf{return}\ true  \label{alg:31}\\}
            
            \Fn{\four{$Target\ list, L_4, L_1$}}{ \label{alg:35}
            \tcp{$Restore\ the\ hotspot\ and\ volume\ status$}
            \If{$L_4-L_1 == Target\ list$}{
            $Send\ FeedbackCmd_2\ and\ ResetCmd$\\
            $Clear\ call\ and\ text\ message\ records$\\
            \label{alg:37}}          
            }   
 \end{algorithm2e}

\section{Metamaterial Design for Covert Long-Range Attack} \label{chap:6}

\subsection{Design Goals and Challenges}\label{chap:6.1}
\cparagraph{Design goal 1} Eliminate audible leakage in the 100 \textit {Hz} - 4000 \textit {Hz} range to maintain the stealthiness of the attack and avoid detection by acoustic leakage.

\cparagraph{Design goal 2} Enhance beamforming to focus inaudible commands more precisely on the target device, enabling longer-range attacks.

To achieve these goals, the system must tackle two key challenges.

\cparagraph{Design challenge 1} How to design an acoustic metamaterial, which ensures that the sound signal is not leaked (for Design Goal 1) but also allows the inaudible sound signal to propagate over a longer distance (for Design Goal 2).

\cparagraph{Design challenge 2} How to achieve a lightweight and portable design that satisfies the stealthy nature of the attack implementation process(for Design Goal 1). This design requires only a minimum number of speakers and an optimal arrangement.

\subsection{Filter Out Audible Leakage}
\label{chap:6.2}
\begin{figure}[!t]
\vspace{-3mm}
\centering
\subfloat[]{
		\includegraphics[scale=0.105]{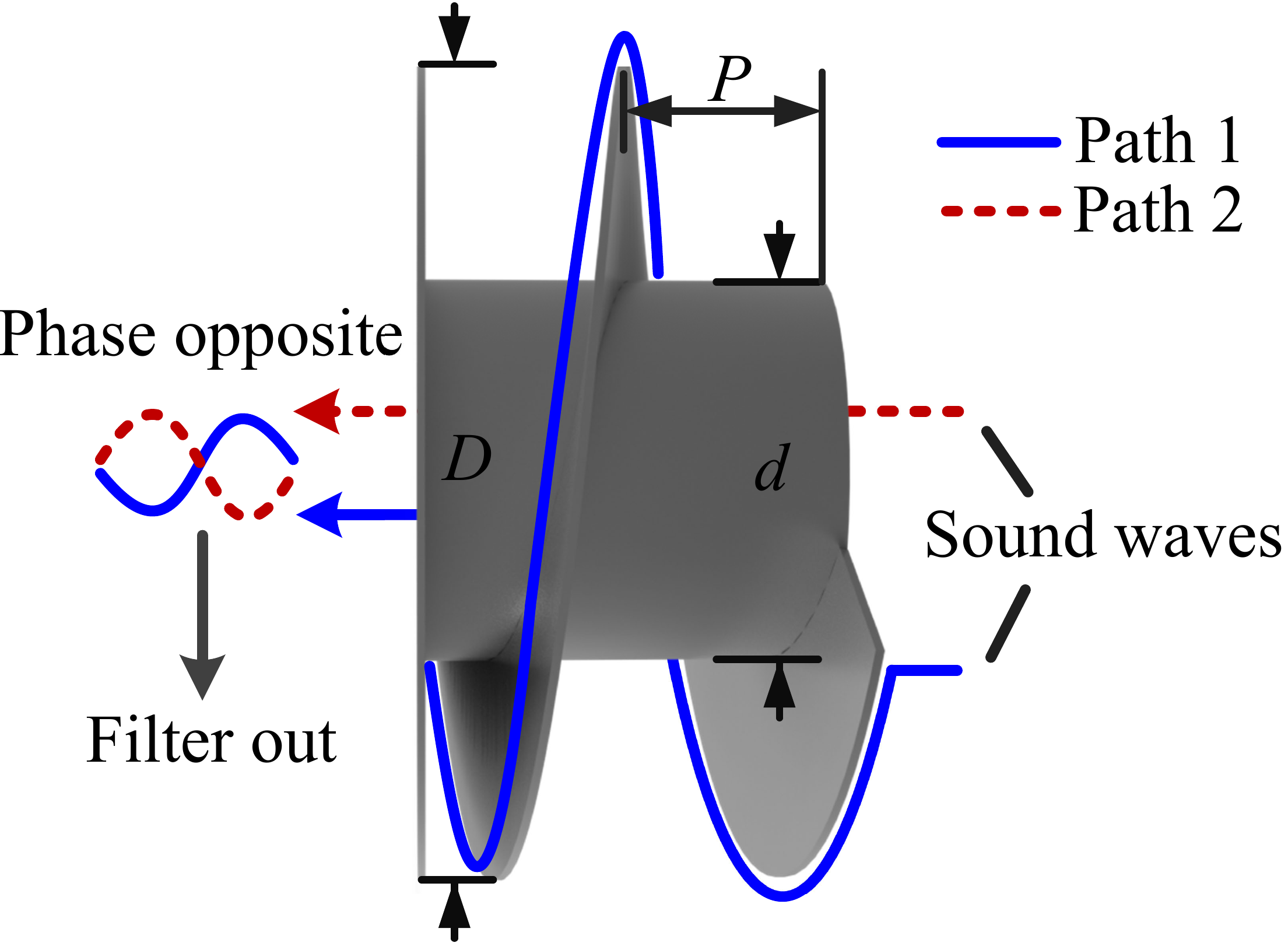}
  \label{filtering principle}}
  \hfill
 % \hfill
\subfloat[]{
		\includegraphics[scale=0.11]{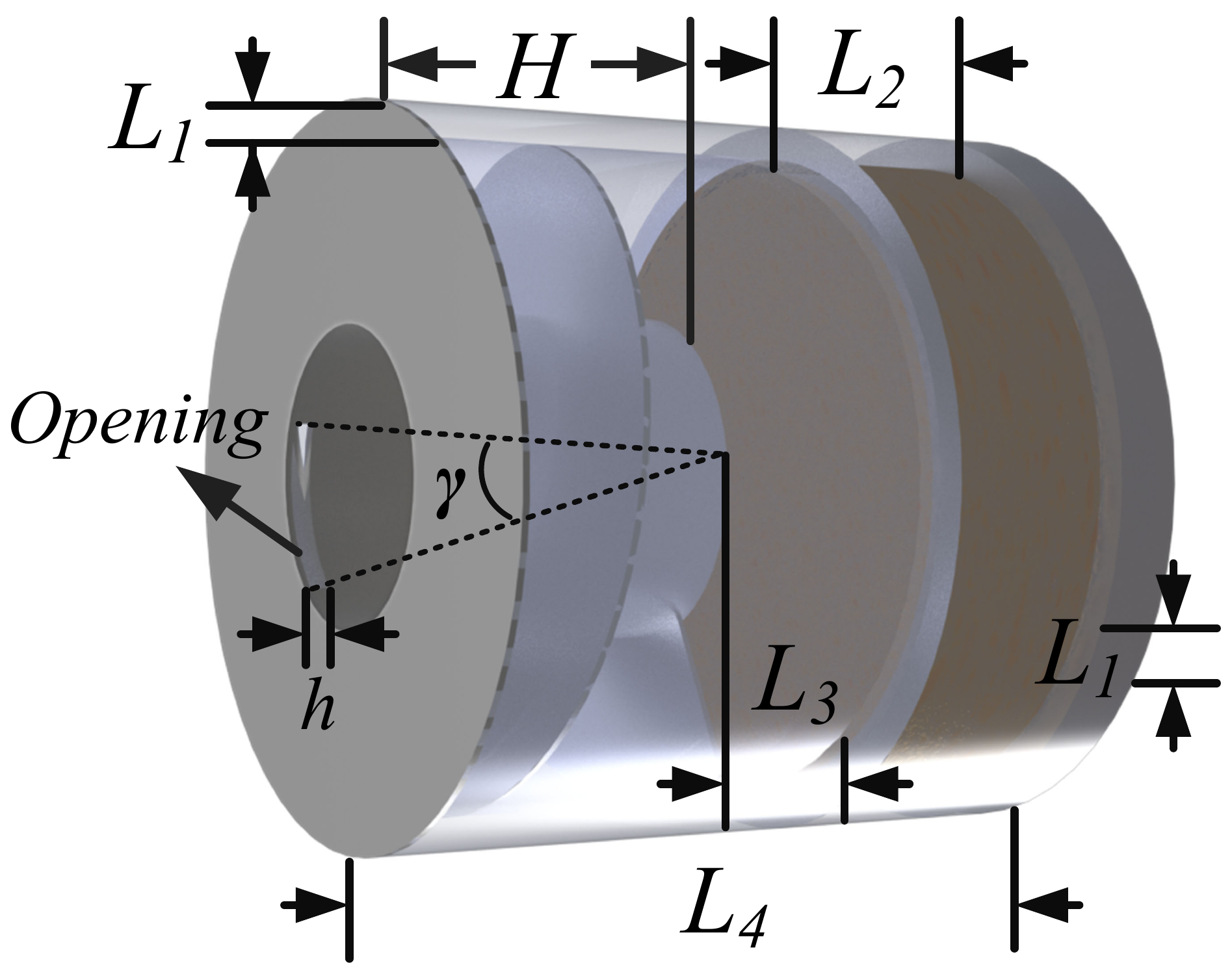}
  \label{LAM}}
\caption{(a) The filtering principle of LAM. (b) LAM- Leak-shielding acoustic metamaterial.}
\label{LAM-A}
\vspace{-15pt}
\end{figure}

\begin{figure}[t!]
\centering
\setlength{\abovecaptionskip}{3pt}
\subfloat[]{
		\includegraphics[scale=0.09]{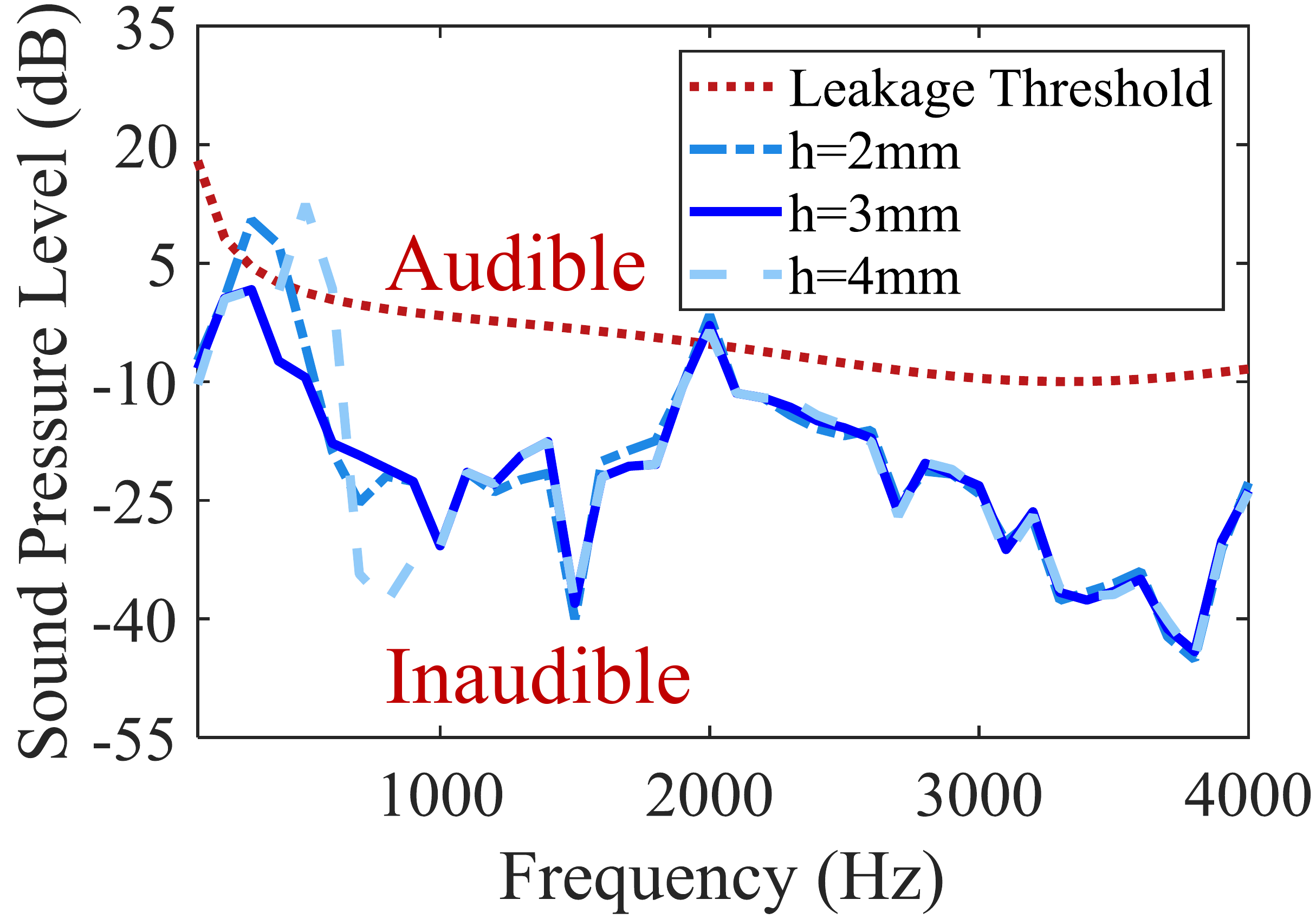}
        \label{highth}}
  \hfill
\subfloat[]{
		\includegraphics[scale=0.09]{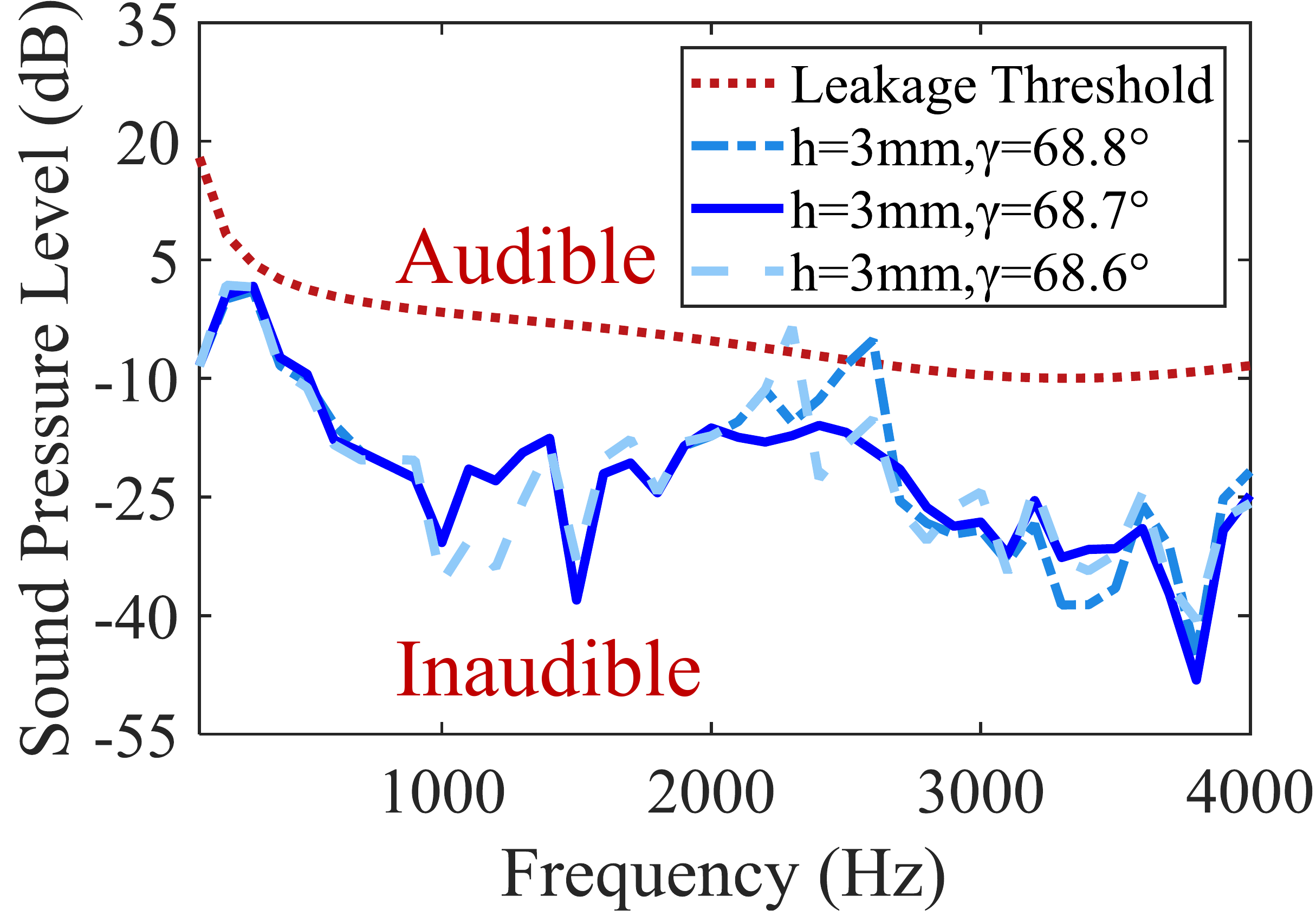}
        \label{angle1}}
\caption{(a) Impact of opening height \( h \) on LAM front filtering. (b) Impact of opening angle \( \gamma \) on LAM front filtering.}
\label{highthandangle}
\vspace{-5pt}
\end{figure}

To address \textit{Challenge 1}, we propose a novel Leak-shielding Acoustic Metamaterial (LAM). The basic filtering mechanism of the LAM is shown in Fig. \ref{filtering principle}. Its internal structure features a circular hole-like design (with a diameter \textit{d} of 22.5 mm) surrounded by a spiral path (with a width \textit{D} of 100 mm and a pitch \textit{P} of 34.542 mm) \cite{r14}. When sound waves of certain frequencies pass through the metamaterial, the path variances induced by its internal structure (e.g., path 1 and path 2) result in phase opposition, leading to filtration by reflection \cite{r31,r32,r33,r95}. However, since the current structure generates a limited number of paths, it can only provide narrowband filtering (600 \textit{Hz} - 1900 \textit{Hz}). This limits its effectiveness in eliminating audible leakage to 100 \textit{Hz} - 4000 \textit{Hz}. To effectively counteract audible leakage, our LAM has undergone significant development:

\cparagraph{Expanding the filtering frequency range to 100 \textit{Hz} - 4000 \textit{Hz}} To further expand the LAM's filtering range, we leverage a key observation in existing works: an opening at the connection between the circular hole-like structure and the surrounding spiral path can expand the LAM's filtering range \cite{r14}. The reason for this observation is that an opening creates more paths, allowing more frequencies to be filtered out through phase opposition. One question that naturally arises is how to determine the opening geometric size. Through our simulations in COMSOL, we found that both the height (\textit{h}) and the angle ($\gamma$) of the opening affect the filtering range of the LAM, with some key data from the simulation results shown in Fig.~\ref{highthandangle}. The \textit{leakage threshold} in Fig.~\ref{highthandangle} represents the sound pressure level (SPL) of audible leakage at different frequencies that can be heard by humans. According to previous studies, such as LipRead~\cite{lipread}, the \textit{leakage threshold} should be 5 dB below the threshold of hearing curve. Therefore, we derive the \textit{leakage threshold} by subtracting 5 dB from the threshold of hearing curve~\cite{lipread}. 

As shown in Fig.~\ref{highth}, the opening height (\textit{h}) significantly affects the filtering performance in the low-frequency range. When $h = 3$ mm, the sound pressure level (SPL) can be effectively suppressed below the leakage threshold. We believe this is because low-frequency sound waves, with their longer wavelengths, tend to propagate through the large-scale internal channels of the structure. In this case, the coupling strength between the central cylindrical structure (determined by \textit{h}) and the rest of the system becomes critical (see Fig.~\ref{LAM}). If \textit{h} is too small, the coupling strength is insufficient, leading to poor filtering performance. If \textit{h} is too large, excessive coupling may occur, resulting in excellent filtering performance at certain frequencies, while the filtering effect at other frequencies fails to effectively suppress leakage. A height of 3 mm strikes a good balance between reflection strength and frequency coverage, ensuring stable low-frequency filtering. In addition, as shown in Fig.~\ref{angle1}, the aperture angle $\gamma$ is similarly crucial for high-frequency filtering. When $\gamma = 68.7^\circ$, the SPL in this range can also be effectively suppressed below the leakage threshold. High-frequency sound waves, with shorter wavelengths, are more prone to strong multipath interference within the spiral blade structure. The angle $\gamma$ directly influences the coupling strength between this structure and the rest of the system (see Fig.~\ref{LAM}). Similar to \textit{h}, a $\gamma$ value that is too small or too large will cause filtering imbalance, whereas $\gamma = 68.7^\circ$ achieves stable filtering in the high-frequency range. Therefore, we find the opening height $h = 3~mm$ and angle $\gamma = {68.7^ \circ }$ can expand the filtering range to 100 \textit{Hz} - 4000 \textit{Hz}.

Although the LAM can filter out audible leakage in front of the loudspeaker, it still struggles to block leakage from the sides or back. To address this, we further optimized its structure:

\cparagraph{Preventing the audible leakage from leaking to other directions} As shown in Fig. \ref{LAM}, first, to prevent audible sound waves from leaking out from the sides of the speaker array, we increase the thickness (${L_1}$) and length (${L_3}$) of the walls on the sides of the LAM to encase the speaker array. This ensures that the audible leakage from the sides is reflected back by the walls. Second, to eliminate the audible leakage reflected to the back, we fill the back of the LAM with melamine sponge (${L_2}$), which absorbs the sound waves through its heat-viscous loss. Notably, we first confirm that the length ${L_3} = 17.5~mm$, which is equal to the length of the speaker array.

\begin{figure}[t!]
\centering
\setlength{\abovecaptionskip}{3pt}
\subfloat[]{
		\includegraphics[scale=0.093]{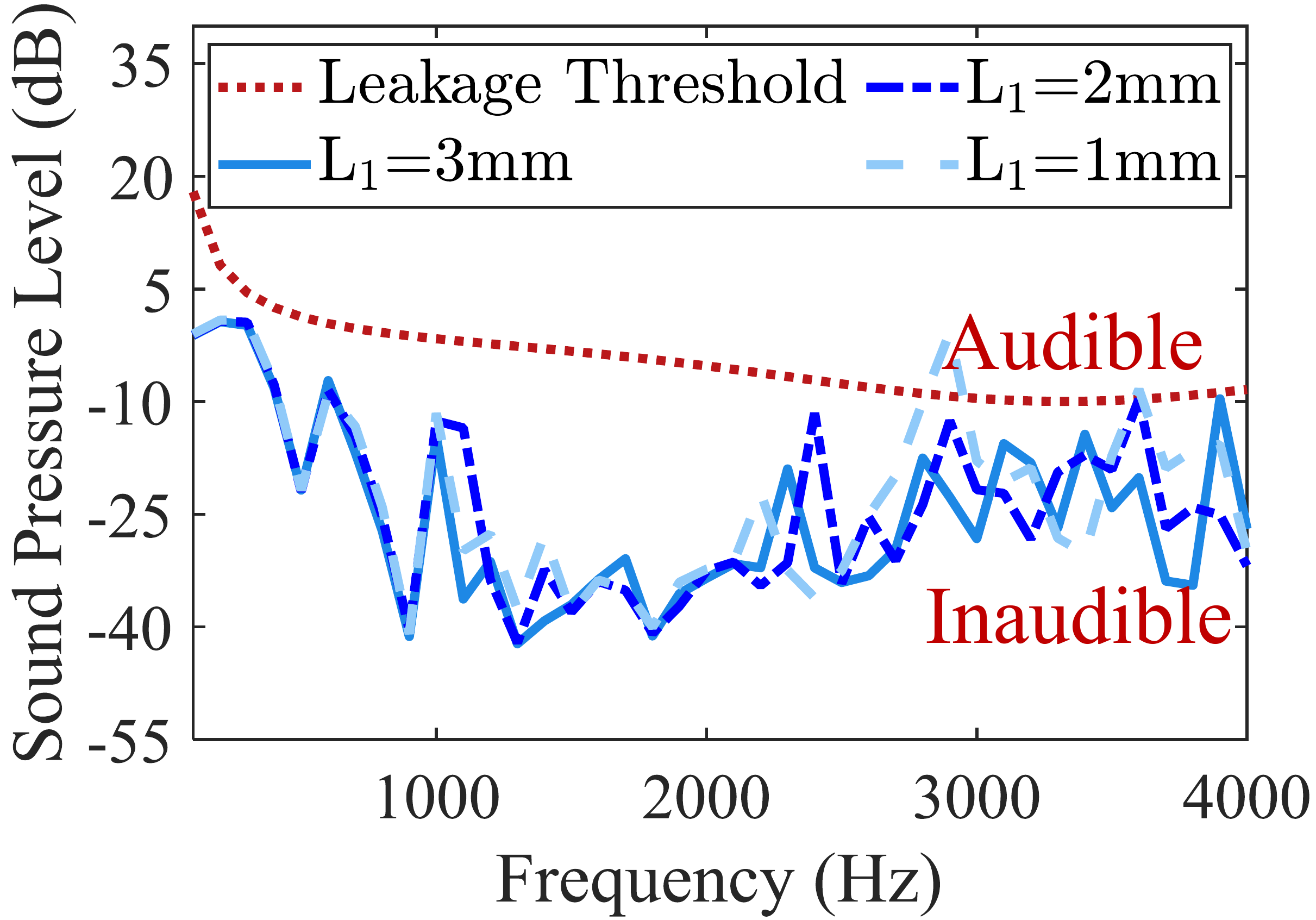}
        \label{sides}}
  \hfill
\subfloat[]{
		\includegraphics[scale=0.093]{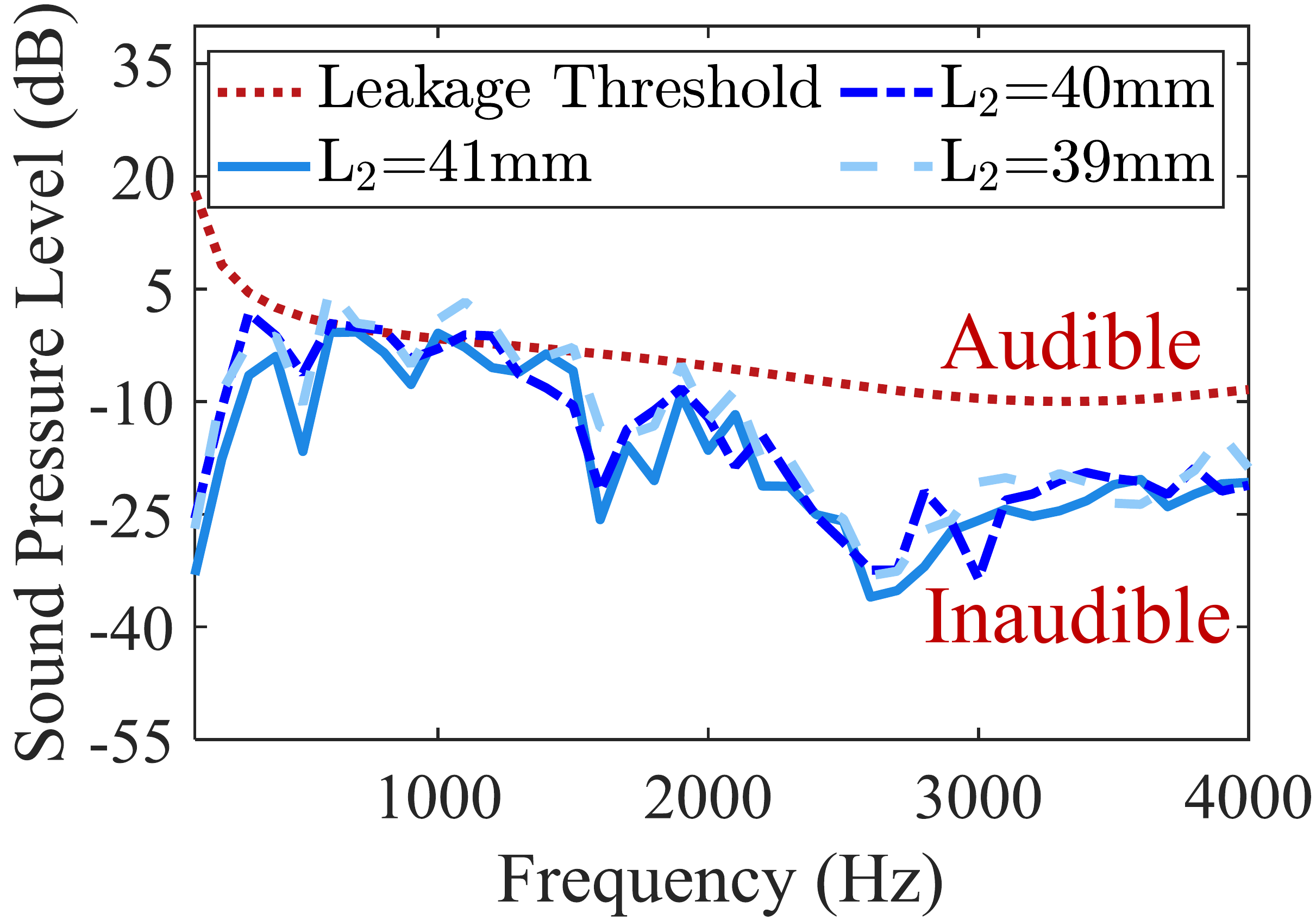}
        \label{back}}
\caption{(a) Impact of sponge thickness on LAM back filtering. (b) Impact of wall thickness on LAM sides filtering.}
\label{Filtering effect}
\vspace{-5pt}
\end{figure}

To investigate the effects of different wall thicknesses $L_1$ and melamine sponge thickness $L_2$ on eliminating sides and back audible leakage, we performed simulations in COMSOL. The simulation results for some representative $L_1$ and $L_2$ in suppressing side and back leakage are shown in Fig.~\ref{sides} and Fig.~\ref{back}. The results in Fig.~\ref{sides} and Fig.~\ref{back} show that when the wall thickness $L_1 > 2~mm$ and the melamine sponge thickness $L_2 > 40~mm$, the audible leakage is below the \textit{leakage threshold}. To ensure inaudibility, it is crucial to choose an appropriate thickness to enhance the compactness of the LAM. Therefore, we determined the wall thickness ${L_1} = 2~mm$ and melamine sponge thickness ${L_2} = 40~mm$, which balance performance and volume, with the total length of the LAM being ${L_4 = H + L_1 + L_2 + L_3} = 109.5~mm$.

Filtering audible leakage against inaudible attacks only addresses the stealth of the attack. Therefore, additional design improvements are required to further enable longer-range attacks with a reduced number of loudspeakers. The key to this advancement is to keep the sound from leaking while ensuring sufficient range.

\subsection{Increase the Range of Inaudible Commands}
\label{chap:6.3}
In the above improvements, the LAM can only play a role in filtering audible leakage in inaudible attacks. However, by carefully designing the shape of incident sound waves, the LAM can be transformed into a piston sound source. The piston sound source can enhance the high-frequency sound waves through acoustic resonance~\cite{r86}, allowing a small loudspeaker array to perform the same beamforming effect as a large loudspeaker array to enhance inaudible commands.
\begin{figure}[t!]
\begin{minipage}[t]{0.48\linewidth}
        \centering        \includegraphics[width=0.95\linewidth]{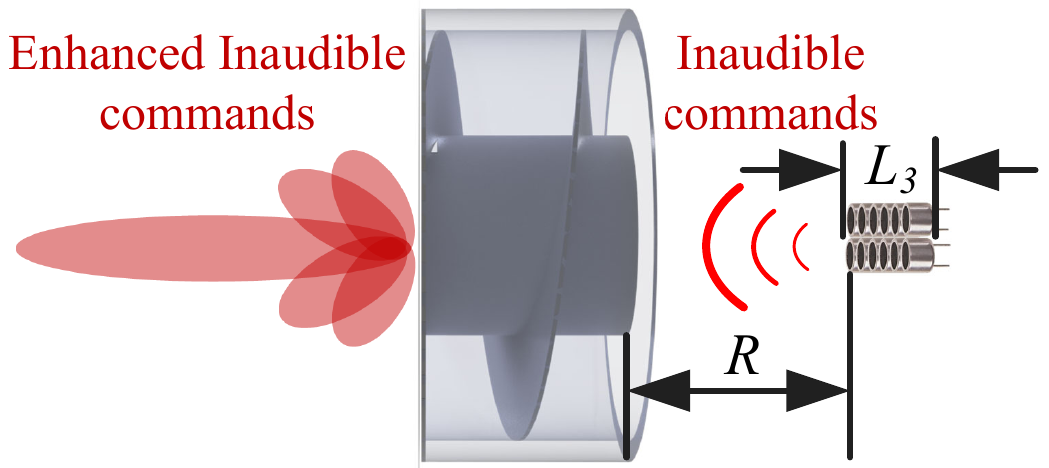}
        \caption{Enhancement effect of inaudible commands in LAM.}
        \label{Enhancement effect of MAM}
    \end{minipage}
    \hfill%\hspace{0.1cm}
\begin{minipage}[t]{0.48\linewidth}
        \centering       
        \includegraphics[width=0.45\linewidth]{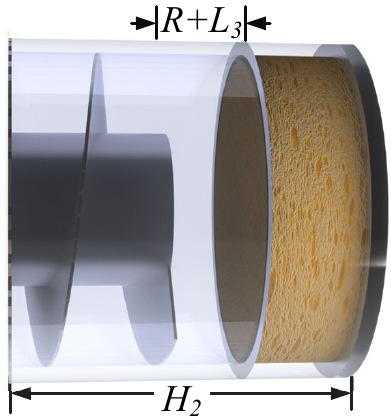}
        \caption{Multi-functional Acoustic Metamaterial (MAM).}
        \label{MAM}
    \end{minipage}
   
 \end{figure}
As shown in Fig.~\ref{Enhancement effect of MAM}, LAM consists mainly of a barrier and a circular piston in the center, and its structure is similar to the piston sound source~\cite{r86,r118}. When the inaudible commands pass through the LAM, the circular barrier causes them to converge into the circular hole-like structure, thus enhancing them. However, from the simulation results, we found that the surrounding spiral path inside LAM would disrupt the incident plane wave of inaudible commands, failing to form a piston sound source (the prerequisite for transforming the LAM to a piston sound source is that the incident waveform of inaudible commands is a plane wave~\cite{r80,r81}).

\begin{table}
    \scriptsize
    \caption{The impact of various factors on the enhancement effect}
    \label{speaker number}
    \vspace{1mm}
    \centering
    \setlength{\tabcolsep}{4.5pt} % 设置列间距为 10pt
    \begin{tabular} {cccc}
    \toprule
    \textbf{Number of Loudspeakers} & \textbf{Range (\textit{R})} &\textbf{Carrier Frequency}&\textbf{Maximum SPL}\\
    \midrule
        2 &  22 mm & 40500 Hz & 128 dB \\
        \rowcolor{gray!20}4 & 14 mm &  40000 Hz & 130 dB\\
        6 &   14 mm    &  40000 Hz&  134 dB\\
        \rowcolor{gray!20}8  &  11 mm & 39500 Hz & 135 dB\\
        10 & 15 mm   &  40500 Hz & 138 dB\\
        \rowcolor{gray!20}\textbf{12} & \textbf{18 mm}   &  \textbf{40200 Hz} & \textbf{142 dB}\\
        14 & 18 mm   &  40200 Hz & 142 dB\\
        \rowcolor{gray!20}16 & 18 mm   &  40200 Hz & 142 dB\\
    \bottomrule
    \end{tabular}
    \vspace{-4mm}
\end{table}

\begin{figure*}[t!]
   \setlength{\abovecaptionskip}{6pt}
    \begin{minipage}[t]{0.23\linewidth}
    \setlength{\abovecaptionskip}{-0.22cm}
        \centering   
        \raisebox{0.8mm} {\includegraphics[width=1\linewidth]{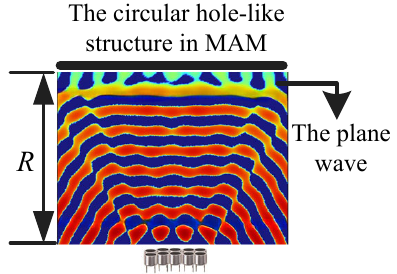}}
        \caption{Speaker array's sound field}
        \label{The sound field}
    \end{minipage}  
    % \hfill%
    \hspace{0.4cm}
    \begin{minipage}[t]{0.21\linewidth}
    \setlength{\abovecaptionskip}{-0.22cm}
        \centering        
        \includegraphics[width=1\linewidth]{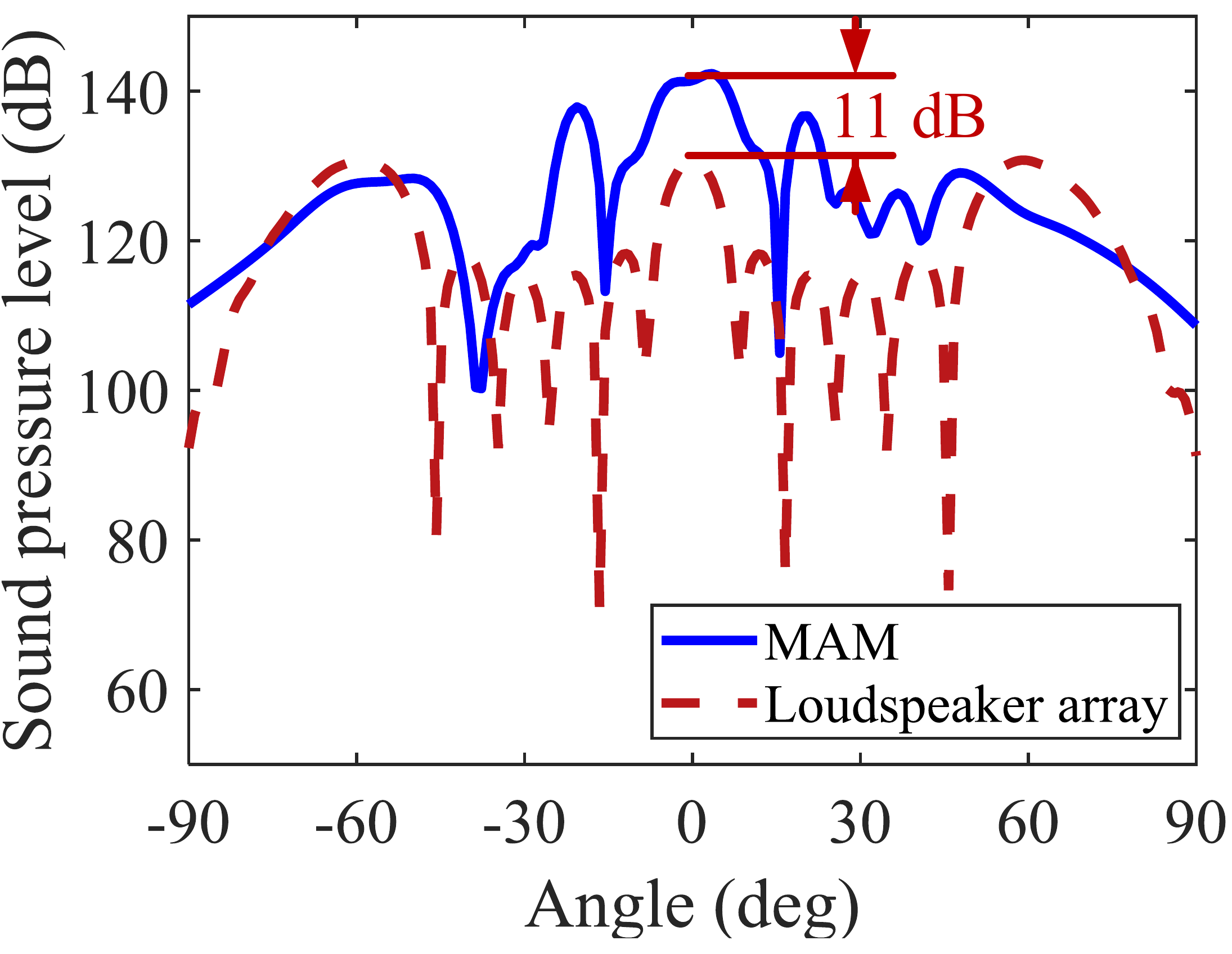}
        \caption{SPL of enhancement.}
        \label{Simulation enhancement effect}
    \end{minipage}
    \hfill
    \begin{minipage}[t]{0.213\linewidth}
    \setlength{\abovecaptionskip}{-0.22cm}
        \centering        
        \raisebox{2mm} {\includegraphics[width=1\linewidth]{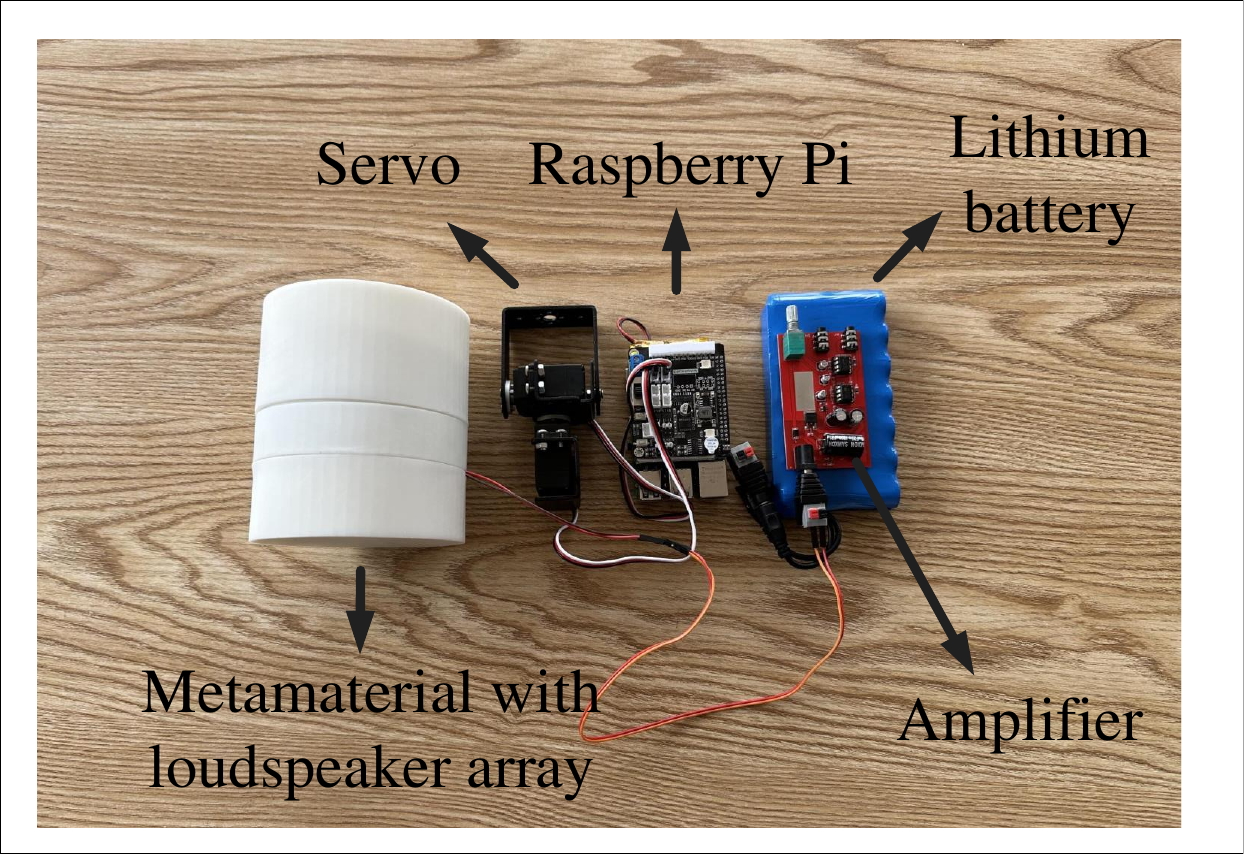}}
        \caption{Prototype of system}
        \label{prototype}
    \end{minipage}
    \hfill
    \begin{minipage}[t]{0.215\linewidth}
    \setlength{\abovecaptionskip}{-0.22cm}
        \centering       
       \raisebox{2mm} {\includegraphics[width=1\linewidth]{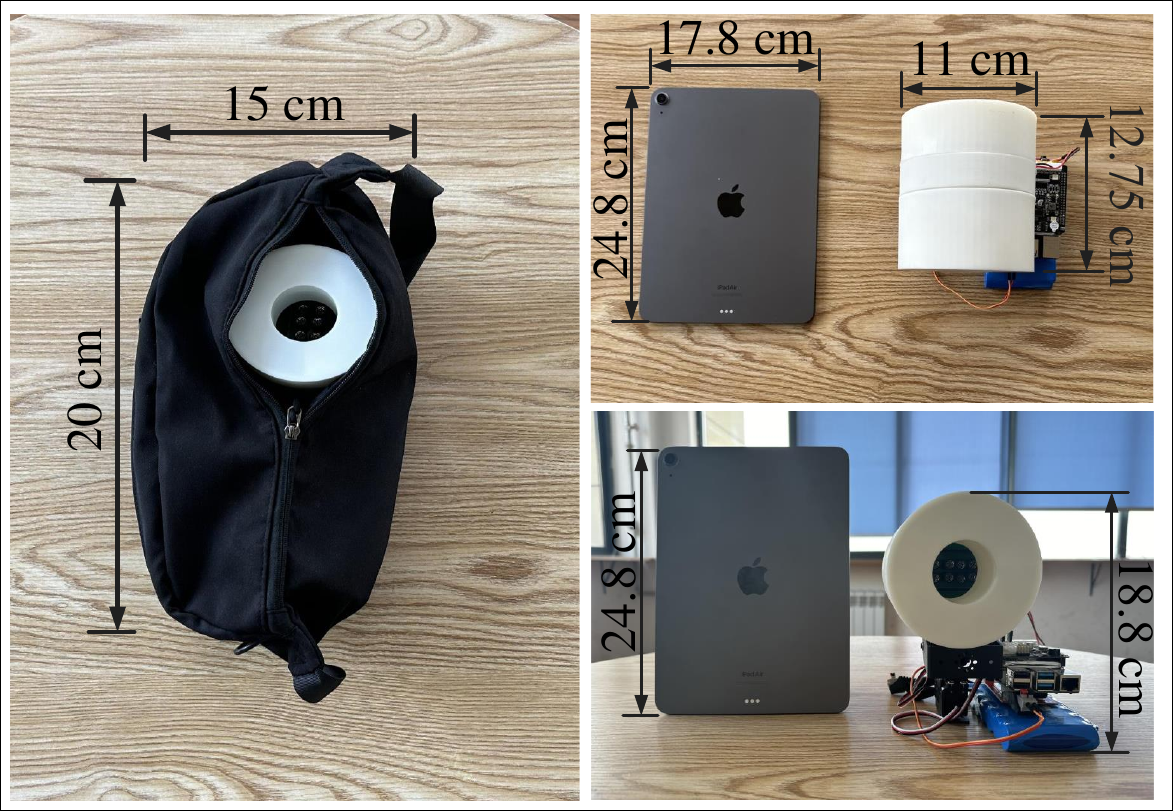}}
        \caption{Covert system}
        \label{covert}
    \end{minipage}  
    \vspace{-10pt}
 \end{figure*}
 
To solve this issue, we let the inaudible commands pass through a LAM under different conditions and observed its waveform. Then, we find that whether the inaudible commands can form plane waves depends on three factors: (1) the number of loudspeakers; (2) the range between the loudspeaker array and circular hole-like structure; (3) carrier frequency of inaudible commands. 
To investigate the impact of these factors, we simulated them in COMSOL. After simulating these factors, we obtained the max SPL under different conditions in Table \ref{speaker number}. We traverse all the possible combinations of the number of speakers, range (\textit{R}), and carrier frequency to investigate their impact. Table \ref{speaker number} provides the optimal combinations of range (\textit{R}) and carrier frequency under different numbers of loudspeakers.
The highest gain is achieved when 12 loudspeakers are incident at 40200 \textit{Hz}, positioned at a range of 18~mm (i.e., $R=18~mm$) from the circular hole-like structure. We also found that when using 14 loudspeakers or more, the SPL no longer increases. This is because the 12 loudspeakers perfectly cover the circular hole-like structure in the LAM, and more loudspeakers will be disturbed by other structures. Furthermore, we conducted a comparative analysis of the various layouts of the 12 speakers in the experiment referenced as Section \ref{A2}. This analysis led to the identification of the optimal configuration, namely, two rows and six columns, which was found to result in the greatest possible attack range. 

The final structure is depicted in Fig. ~\ref{MAM}. This final structure is referred to as the Multi-functional Acoustic Metamaterial (MAM). Its length is ${H_2}= {L_4} + R  = 127.5~mm$. To verify its performance, we first simulate the sound field of the loudspeaker array. The result is shown in Fig. \ref{The sound field}. We have proven that under this structure, the waveform of the loudspeakers is a plane wave when it reaches the circular hole-like structure.
We then investigate its enhancement effect in COMSOL. As shown in Fig. \ref{Simulation enhancement effect}, we find that when inaudible commands meet the above conditions and pass through the MAM, the SPL increases by 11~dB, indicating that the energy of inaudible commands has been enhanced. 

\section{IMPLEMENTATION} \label{chap:7}
\ifx\allfiles\undefined

\else
\fi

\subsection{\SystemName Components} 

The prototype of \SystemName uses commercial devices and a 3D-printed acoustic metamaterial (MAM), as shown in Fig. \ref{prototype}. This metamaterial filters out low-frequency sounds and amplifies inaudible commands. The Raspberry Pi serves as the terminal, responsible for sending attack commands, controlling the servo motor, and running the feedback mechanism. The speaker array(12 cm × 1.4 cm) in the MAM transmits commands with a center frequency of 40 kHz.

\subsection{Stealthiness and Portability} The stealthiness and portability of inaudible attacks are crucial for covert operations and evading detection. As shown in Fig. \ref{covert}, the dimensions of the assembled \SystemName are 12.75 cm × 11 cm × 18.8 cm, making it compact and easy to carry. To enhance its stealth capabilities, we have consolidated all the attack devices into a travel bag that can be held by hand or slung over the shoulder during silent attacks. This compact and discreet design makes \SystemName in its travel bag configuration an effective tool for silent operations.

\ifx\allfiles\undefined

\else
\fi
 \section{EVALUATION} \label{chap:8}
\subsection{Experimental Setup}

\cparagraph{Target devices} We evaluated four mainstreamed voice-controlled personal assistants on smartphones and smart devices, including Apple Siri \cite{Applesiri}, Amazon Alexa \cite{AmazonAlexa}, Google Assistant \cite{GoogleAssistant}, Xiaomi Xiaoai \cite{XiaomiXiaoai} and Huawei Xiaoyi \cite{HuaweiXiaoyi}. In our evaluation, we use twelve devices and the specifications of the devices are given in the Table \ref{target devices}.

\begin{table}[t!]
    \scriptsize
    \caption{Target devices}
    \label{target devices}
    \vspace{1mm}
    \centering
    \setlength{\tabcolsep}{8.5pt} % 设置列间距为 10pt
    \begin{tabular} {p{1cm}ccccc}
    \toprule
    \textbf{Manuf.} & \textbf{Model} & \textbf{OS} & \textbf{Voice system}\\
    \midrule
    \rowcolor{gray!20}  & iPhone 14 Pro & IOS 17.4.1 & Siri \\
    \rowcolor{gray!20}  & iPhone 13 Pro Max & IOS 17.2.1 & Siri \\
    \rowcolor{gray!20}  & iPhone 13 &  IOS 17.2.1 & Siri\\
    \rowcolor{gray!20} \multirowcell{-4}{Apple} & iPhone 12 &  IOS 16.5.1 &  Siri\\
    
    \multirowcell{1}{\makecell[l]{~Amazon}} &  Echo Dot 5th     & 9698496900h & Alexa \\
    
    \rowcolor{gray!20} \makecell[l]{~Samsung} &  Galaxy S7  & Android 8.0.0 & Google Assistant \\
    
    \multirowcell{-0.5}{~Google} & Pixel 3 &  Android 12.0.0 &  Google Assistant\\
    &  Pixel 2  &  Android 10.0.0 &  Google Assistant\\
   
    \rowcolor{gray!20} &  Redmi K50 Ultra  &  Miui 13.0.10 &  XiaoAI\\
    \rowcolor{gray!20} \multirowcell{-1}{~Xiaomi} &Xiaoai Play 2 &  1.62.26 &  XiaoAI\\
    \rowcolor{gray!20} &  Mi10 Lite Zoom  &  Miui 12.0.6 &  XiaoAI\\
     
    \multirowcell{1}{~Huawei} &  nova 7 SE 5G  & HarmonyOS 3.0.0 & Xiaoyi \\
   
    \bottomrule
    \end{tabular}
\vspace{-10pt}
\end{table}

\begin{figure*}[!t]
\centering
\setlength{\abovecaptionskip}{0pt}
\subfloat[S1]{
		\includegraphics[scale=0.384]{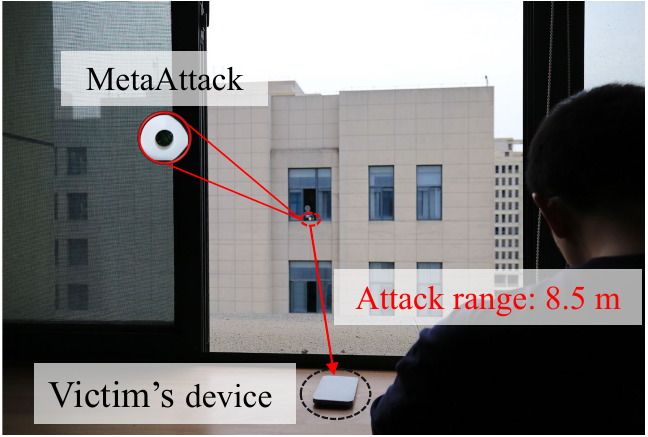}
  \label{S1}}
\subfloat[S2]{
		\includegraphics[scale=0.438]{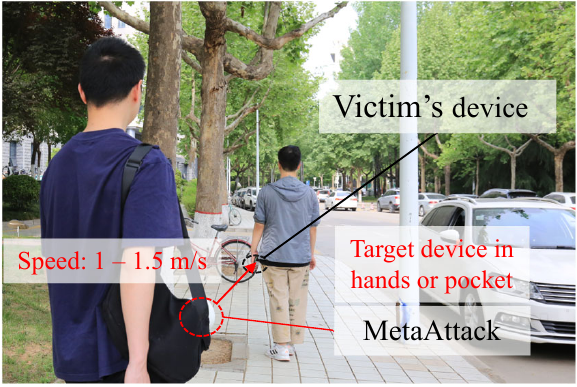}
  \label{S2}}
\subfloat[S3]{
		\includegraphics[scale=0.385]{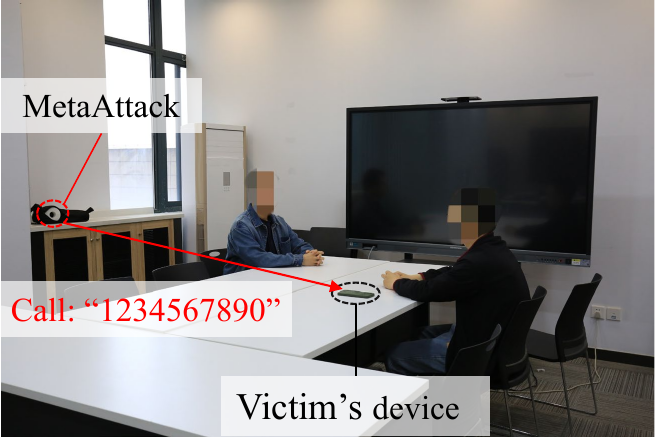}
  \label{S3}}
\subfloat[S4]{
		\includegraphics[scale=0.386]{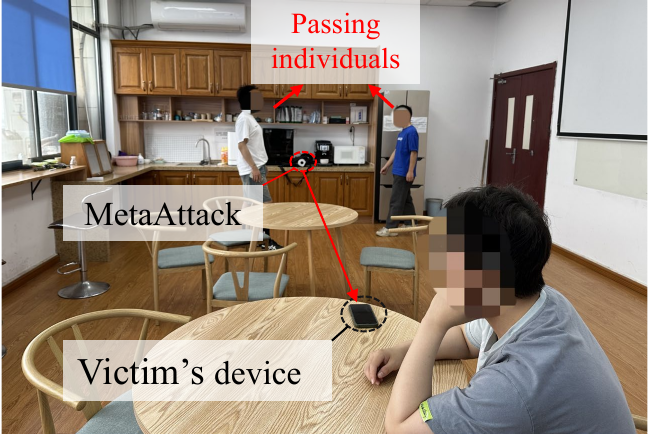}
  \label{S4}}
\caption{Real attack scenarios are as follows: (a) S1: The attacker launches a remote attack from a building 8.5 meters away; (b) S2: The attacker targets a walking victim outdoors; (c) S3: The attacker sends a call command to eavesdrop; (d) S4: The attacker operates in an environment with obstacles.}
\label{scenarios}
\end{figure*}

\cparagraph{Metrics} Table \ref{Evaluating metrics} lists the quantified metrics used in our evaluation. The success rate is a key metric for evaluating inaudible attacks. We measure the attack range by checking whether the attack system can achieve a success rate of 50\% or more within the range (RSA). Word accuracy measures the ratio at which inaudible commands are correctly recognized. Precision and Recall measure the accuracy of the \SystemName feedback system.

\cparagraph{Experiment design} We conducted a range of experiments, all of which, except experiment S3, used ``Send a message to Jack'' as the inaudible command to achieve the purpose of sending false information. Experiments A1 to A5 evaluated the basic performance of \SystemName and verified the contribution of the metamaterials. The basic performance tests include attack range (A1), selection of loudspeaker array arrangement (A2), and the impact of environmental noise (A3); the contribution of the metamaterials is assessed through their ability to suppress audible leakage (A4) and the performance improvements brought by MAM and LAM (A5). Experiments S1 to S5 evaluated the attack performance and the accuracy of the feedback system. As shown in Fig. \ref{scenarios}, the attack distances for scenarios S1 to S4 all exceed 4 meters and are located in outdoor or small private spaces, making them unsuitable for large external devices. Existing methods typically involve a trade-off between stealthiness and attack range~\cite{r7,r9,perturbation,dol}, whereas \SystemName performs better in this regard. Detailed experimental settings can be found in Table \ref{experimental setting}.

 \begin{table}[t!]
    \scriptsize
    \caption{Evaluating metrics}
    \label{Evaluating metrics}
    \vspace{1mm}
    \centering
     \setlength{\tabcolsep}{4pt}% 设置列间距为 10pt
    % \hfill
    \begin{tabular} {p{1.5cm}p{6.8cm}}
    \toprule
    \multicolumn{1}{c}{\textbf{Metrics}} & \multicolumn{1}{c}{\textbf{Description}}\\
    \midrule
        \rowcolor{gray!20} \multirowcell{1.7}{Success rate}  & Number of successful attacks / Total attempts (30 trials per device per round)~\cite{r100,r9}. \textit{For Exp. A1(\ref{A1}).} \\
        
        \multirowcell{2.5}{RSA} & The \textit{Range} at which \SystemName can \textit{Successfully Attack} (with a success rate of over 50\% )~\cite{lipread}. \textit{For Exps. A2(\ref{A2}), A4(\ref{A4}), A5(\ref{A5}), S1(\ref{S-1}), S2(\ref{S-2}), S4(\ref{S-4}).}  \\
        
        \rowcolor{gray!20}\multirowcell{1.7}{Word accuracy} & Proportion of correctly recognized words in inaudible commands~\cite{dol,lipread,longdol}. \textit{For Exps. S3(\ref{S-3}).}  \\

        \multirowcell{2.7}{Precision} & The proportion of successful attacks correctly identified to all successful attacks identified by the feedback mechanism~\cite{eararray,normal,lipread}. \textit{For Exp. S5(\ref{S5}).} \\
        
        \rowcolor{gray!20} \multirowcell{1.6}{Recall} &  The proportion of successful attacks correctly identified to all truly successful attacks~\cite{eararray,normal,lipread}. \textit{For Exp. S5(\ref{S5}).}\\
    \bottomrule
    \end{tabular}
    \vspace{-5pt}
\end{table}

\begin{table*}
    \scriptsize
    \caption{List of experimental setting}
    \label{experimental setting}
    \vspace{1mm}
    \centering
    \begin{tabular} {p{1.6cm}p{1.6cm}p{3.6cm}p{9.5cm}}    \toprule
    \multicolumn{1}{c}{\textbf{Test objectives}}  & \multicolumn{1}{c}{\textbf{Label}} &\multicolumn{1}{c}{\textbf{Test focus}} & \multicolumn{1}{c}{\textbf{Description}}\\
    \midrule
\rowcolor{gray!20}  & A1 (Sec.\ref{A1})&  Attack range performance & Attack range of \SystemName was evaluated in comparison to a standalone speaker array.\\  

\rowcolor{gray!20}& A2 (Sec.\ref{A2})& Differences in speakers' layout & Six common array layouts were tested to identify the optimal configuration (Sec.~\ref{chap:6.3}).\\ 

\rowcolor{gray!20}  & A3 (Sec.\ref{A3}) & Noise impact& \SystemName's performance was tested by varying background noise from 55 to 75 dB.\\ 

\rowcolor{gray!20}& A4 (Sec.\ref{A4}) & Leakage filtering performance & The audible sound leakage of \SystemName was measured across different directions from 100-4000 Hz, alongside tests on human perception.\\

\rowcolor{gray!20}\multirowcell{-4}{ \makecell[l]{Basic \\performance \\evaluation}}& A5 (Sec.\ref{A5})& Metamaterials' enhancement & Enhancement performance was evaluated by comparing MAM, LAM, and speaker array.\\

\multirowcell{3.3}{\makecell[l]{Real-world\\ performance\\ evaluation}} & S1 (Sec.\ref{S-1}) &  \makecell[l]{Building-to-building attack}& Target is attacked from 8.5 m away in a building under varying wind speeds (Fig. \ref{S1}).\\ 
        
& S2 (Sec.\ref{S-2}) &  Attack against outdoor targets & The target walking outdoors was attacked, while also verifying the attack effectiveness with the device in a pocket (Fig. \ref{S2}).\\ 
        
& S3 (Sec.\ref{S-3}) & \makecell[l]{Eavesdropping attacks}& \SystemName's eavesdropping was assessed by triggering a call (Fig. \ref{S3}).\\ 
        
& S4 (Sec.\ref{S-4}) &  \makecell[l]{Impact of passing individuals}& Attacks were conducted in a room with passing individuals to explore their impact (Fig. \ref{S4}).\\

& S5 (Sec.\ref{S5}) &  Feedback mechanism performance & Measured the precision and recall of the feedback mechanism in S1-S4 scenarios.\\

\rowcolor{gray!20}  \multirowcell{1}{\makecell[l]{Compared to\\ prior research}} & \makecell[l]{Sec. \ref{N3}} &  \SystemName's concealment advantages & \SystemName's device size and range was benchmarked against prior systems to assess its enhanced covert capabilities.\\                             

    \bottomrule
    \end{tabular}
    \vspace{-2mm}
\end{table*}

\subsection{Performance under Various Conditions }

\subsubsection{A1 - Attack range performance between \SystemName and loudspeaker array\label{A1}} Attack range is a key factor in evaluating the practicality of ultrasonic attacks—greater range offers better concealment. Device microphones and voice systems also influence \SystemName's performance.

\begin{figure} [t!]
\setlength{\abovecaptionskip}{-5pt}
    \centering
    \includegraphics[width=0.95\linewidth]{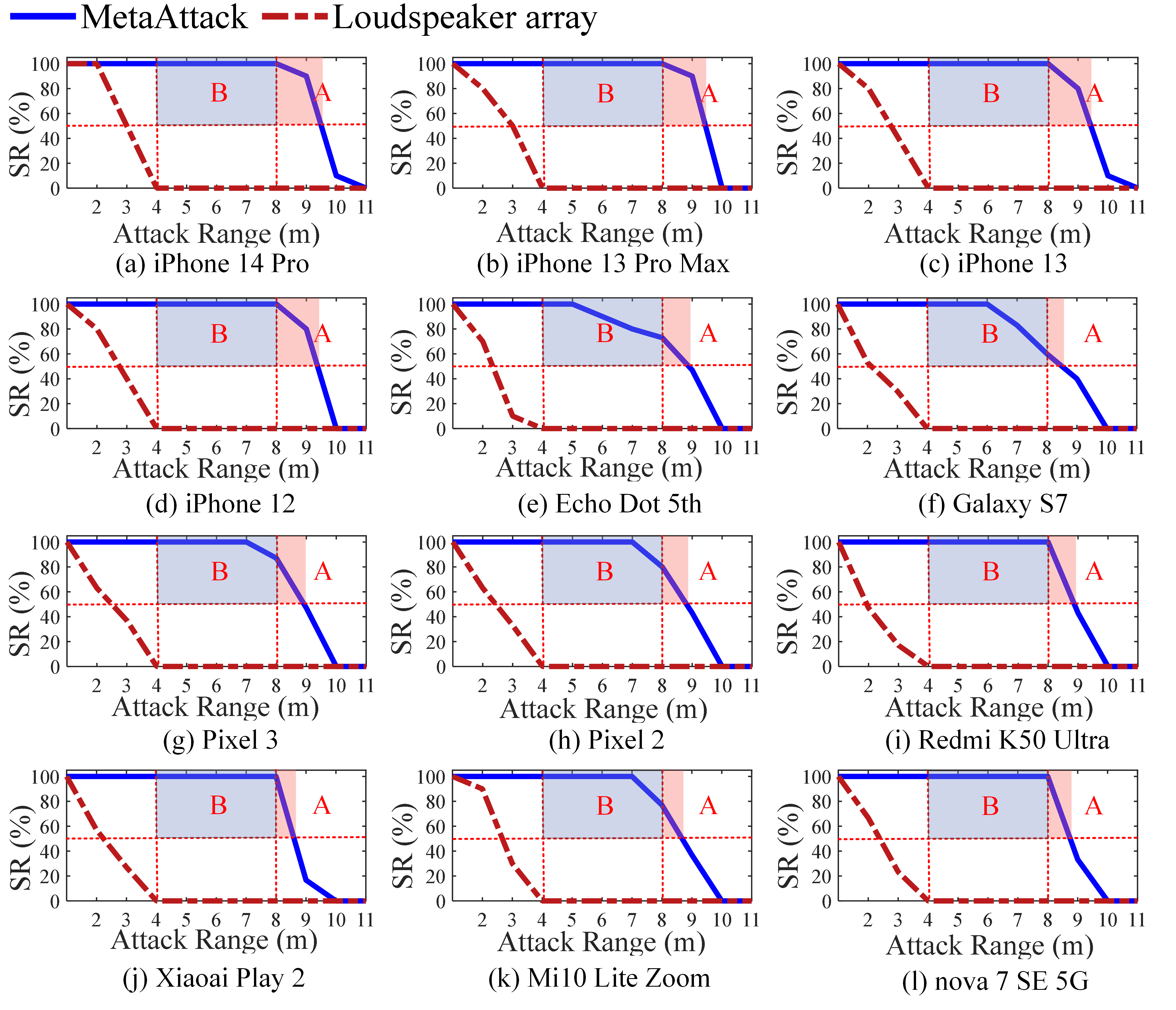}
    \caption{Success rate(SR) performance of \SystemName on various devices, Part A and Part B refer to the quadrants in Fig~\ref{Inaudible research}.}
    \label{devices and range}
     \vspace{-15pt}
\end{figure}

We tested the maximum attack range of \SystemName across different devices (results shown in Fig. \ref{devices and range}). The iPhone 14 Pro performed best, achieving a 90\% success rate at a distance of 9.2 meters. Other devices had an average range of 8.85 meters with success rates above 76\%. \SystemName successfully covered Regions A and B in Fig. \ref{Inaudible research} on all devices, while a regular speaker array had an average range of only 2.72 meters. The performance improvement is due to the MAM's filtering and amplification capabilities, which significantly enhance signal strength.

\subsubsection{A2 - Differences in speaker's layout\label{A2}} 
\begin{figure}[t!]
\centering
\subfloat[]{
\includegraphics[scale=0.53]{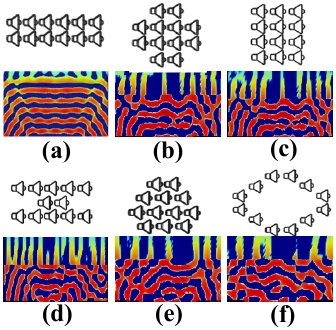}
 \label{newarrange1-1}}
\hfill
\subfloat[]{
\includegraphics[scale=0.49]{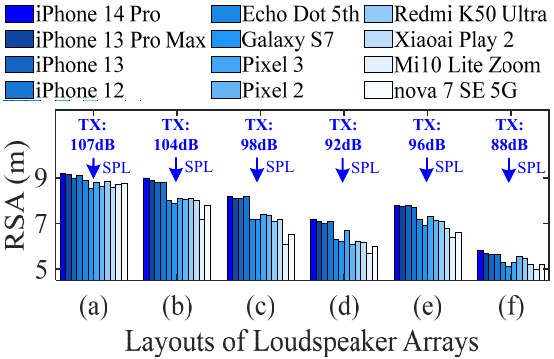}
 \label{newarrange1-2}}
\caption{(a) Six speaker arrays with their sound fields. (b) Performance of speaker arrays (tests include arrays' RSA and SPL at Transmitter(TX).}
\label{arr}
    \vspace{-10pt}
\end{figure}

Our simulation results show that the beamforming effect of \SystemName is most significant when using 12 loudspeakers (see Section \ref{chap:6.3}). However, we found that the speaker layout significantly impacts \SystemName's performance. To determine the optimal speaker layout, we simulated the sound pressure fields of six common configurations in COMSOL and systematically evaluated their RSA and SPL performance, as shown in Fig.~\ref{arr}. As illustrated in Fig.~\ref{newarrange1-1}, only layout (a) successfully forms an effective plane wave, while the others fail to do so. We attribute this to the fact that in layout (a), all speakers are positioned within the central circular opening of the metamaterial, avoiding interference from the surrounding spiral structures. Additionally, the speakers are evenly spaced and symmetrically arranged, which promotes coherent wavefront formation and facilitates the generation of a desirable plane wave. In contrast, the other layouts either suffer from obstruction by the spiral structures or lack consistent spacing and symmetry, which hinders effective wavefront shaping. Fig.~\ref{newarrange1-2} further confirms that layout (a) achieves the best performance in both SPL and RSA, leading us to select it as the final design for the system.

\begin{figure}[t!]
\centering
\subfloat[]{
\includegraphics[scale=0.25]{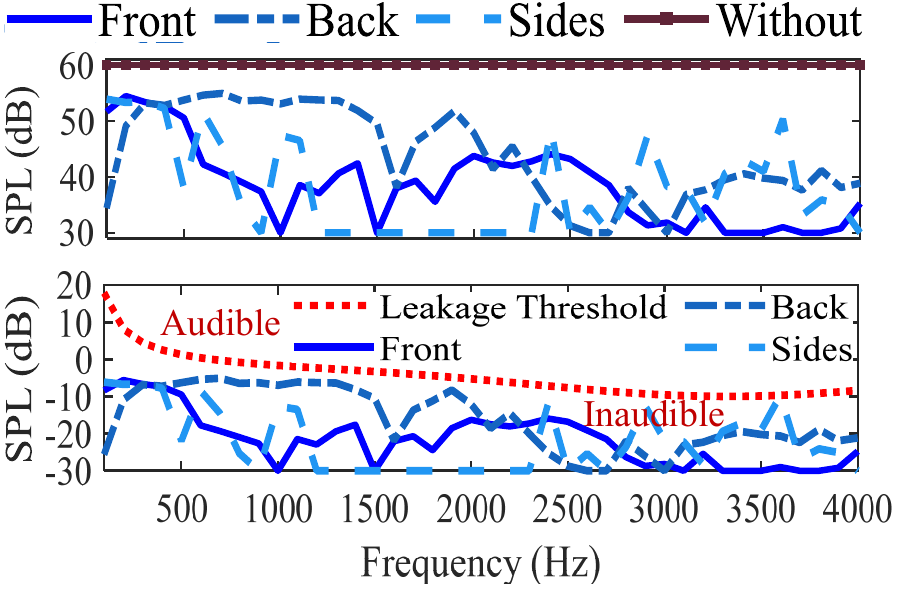}
 \label{newl}}
\hspace{0.1cm}
\subfloat[]{
\includegraphics[scale=0.295]{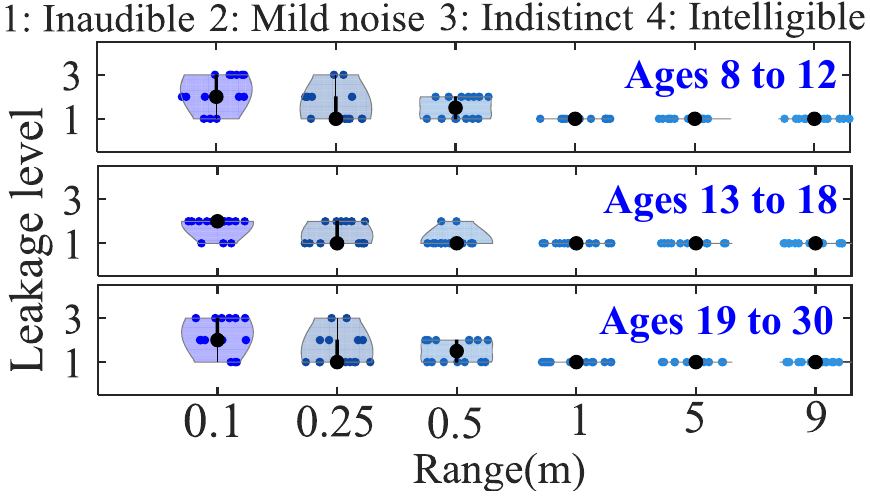}
 \label{Volunteer}}
\caption{(a) Leakage SPL comparison at TX (top) and its relation to the \textit{leakage threshold} (bottom). (b) Volunteers’ perception of the leakage.}
\label{Filtering performance}
    \vspace{-10pt}
\end{figure}

\begin{figure}[t!]
\centering
\subfloat[]{
\includegraphics[scale=0.27]{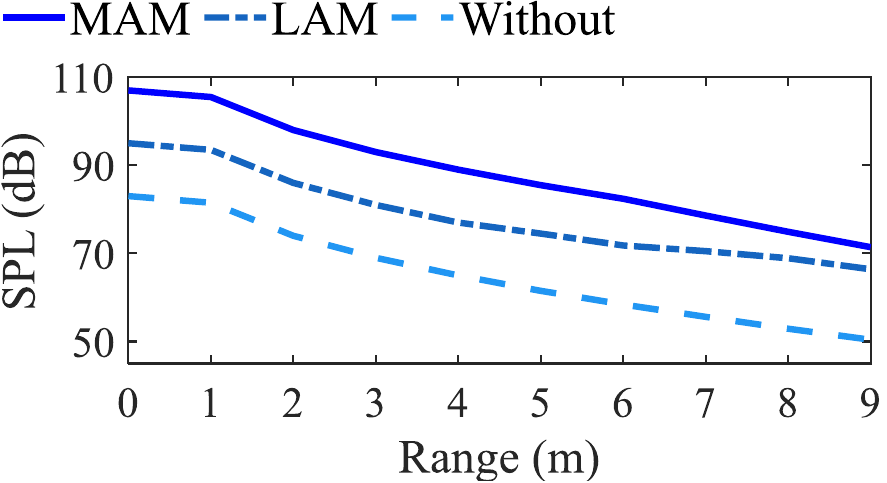}
 \label{ENSPL}}
\hfill
\subfloat[]{
\includegraphics[scale=0.27]{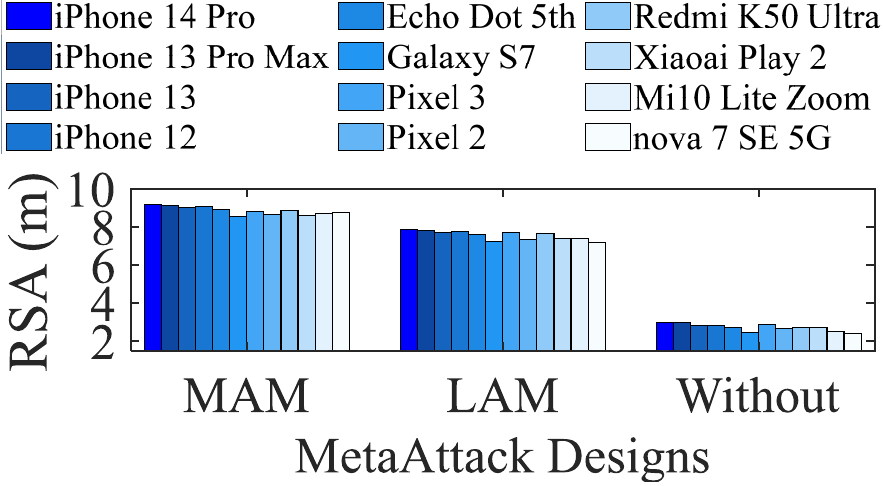}
 \label{ENRANGE}}
    \caption{(a) SPL comparison of inaudible attacks with and without MAM. (b) RSA performance comparison.}
\label{fig: 17}
    \vspace{-10pt}
\end{figure}

\subsubsection{A3 - Noise impact\label{A3}} Another factor affecting the success rate of inaudible attacks is environmental noise, which can interfere with the accuracy of command recognition. We tested \SystemName's RSA on 12 devices under varying noise conditions. The results, shown in Fig. \ref{noise}, indicate that only when the noise level increases to 70-75 dB (similar to noisy office or urban traffic sounds) does the attack range decrease to 7-8 meters (a reduction of about 1 meter). This suggests that the strong beamforming effect of \SystemName allows it to successfully perform inaudible attacks even in typical noisy environments.

\subsubsection{A4 - Leakage filtering performance\label{A4}} To ensure the stealth of the attack, \SystemName needs to effectively filter out audible leakage. In a laboratory environment with a background noise level of 30 dB, we played single-tone signals ranging from 100 to 4000 Hz at 60 dB SPL and used the as-k6 noise tester to evaluate the filtering performance of \SystemName. As shown in Fig. \ref{newl}, \SystemName significantly attenuates sound waves from the front, sides, and rear, reducing audible leakage to below the \textit{Leakage Threshold}.

To verify whether the sound filtered by \SystemName is imperceptible to the human ear, we invited 42 volunteers to participate in the experiment on a voluntary basis. The volunteers were evenly divided into three age groups (14 volunteers per group): 8–12 years old (children, more sensitive to sound), 13–18 years old (teenagers, broader hearing range), and 19–30 years old (young adults, peak auditory resolution), with a balanced gender distribution within each group. The experiment was conducted in an environment with a background noise level of 30~dB. Volunteers stood at various distances and attempted to identify whether there was any audible leakage at four levels: inaudible, mild noise, indistinct and intelligible~\cite{lipread,dol,longdol,2014inaudible}.

As shown in Fig. \ref{Volunteer}, the experimental results indicate that only within a short distance of 0.1 to 0.5 meters could a few volunteers with more sensitive hearing perceive faint or indistinct sounds. We speculate that this may be due to slight audible leakage caused by nonlinear acoustic effects during air transmission.
However, our attack range far exceeds this distance (reaching up to 8.85 meters), and the background noise level in the experimental environment was only 30 dB, which is extremely quiet. In real-world environments, such weak audio leakage is typically masked by higher background noise. Therefore, we believe that \SystemName demonstrates strong imperceptibility in practical attack scenarios.

\subsubsection{A5 - Contribution of metamaterials \label{A5}} The enhancement functionality of MAM plays a key role in enabling 12 compact loudspeakers to surpass large speaker arrays in achieving long-range inaudible command transmission \cite{lipread}. Through experiments conducted in real-world scenarios using ablation techniques, we compared the RSA of MAM, LAM and speaker array without metamaterial on 12 distinct target devices, and analyzed the SPL at various distances.

As shown in Fig. \ref{ENSPL}, the results show that MAM (8.85~m) improves the average attack distance by 14.2\% compared to LAM (7.75~m) and by 225\% compared to the speaker array (2.72~m). Compared to LAM (95~dB), MAM (107~dB) increases the output SPL by 12.6\%, and by 28.9\% compared to the speaker array (83~dB). These results highlight the two key functions of MAM—audible leakage suppression and energy amplification—as essential to enhancing speaker array performance.

\subsection{Performance in Real-world Scenarios}

 \begin{figure}[t!]
\begin{minipage}[t]{0.47\linewidth}
\setlength{\abovecaptionskip}{-0.6cm}
        \centering        
        \raisebox{3.1mm}{\includegraphics[width=1\linewidth]{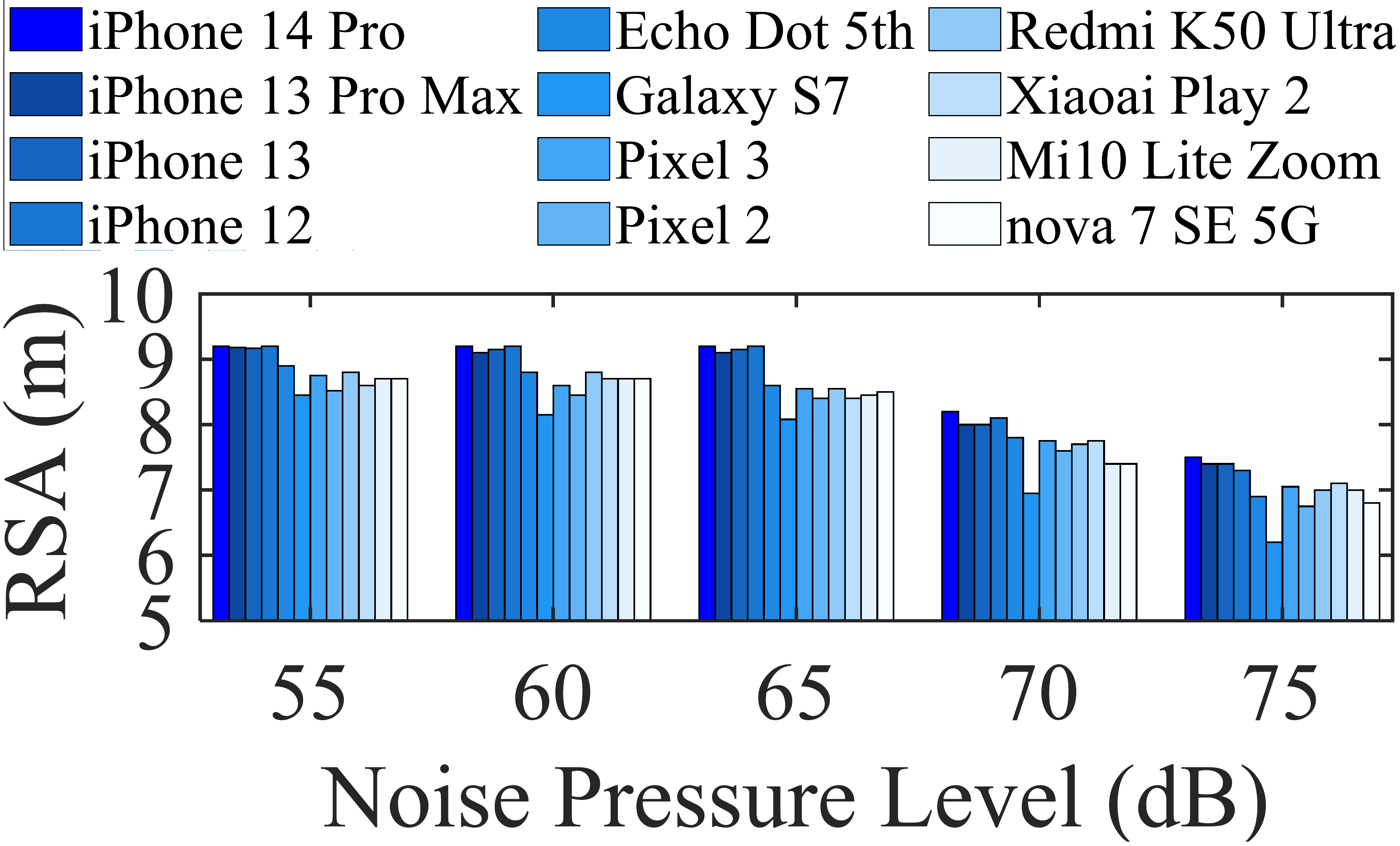}}
        \caption{Impact of noise levels.}
        \label{noise}
    \end{minipage}
            \hfill
        \begin{minipage}[t]{0.47\linewidth}
        \setlength{\abovecaptionskip}{-0.6cm}
        \centering
        \raisebox{3.1mm}{\includegraphics[width=1\linewidth]{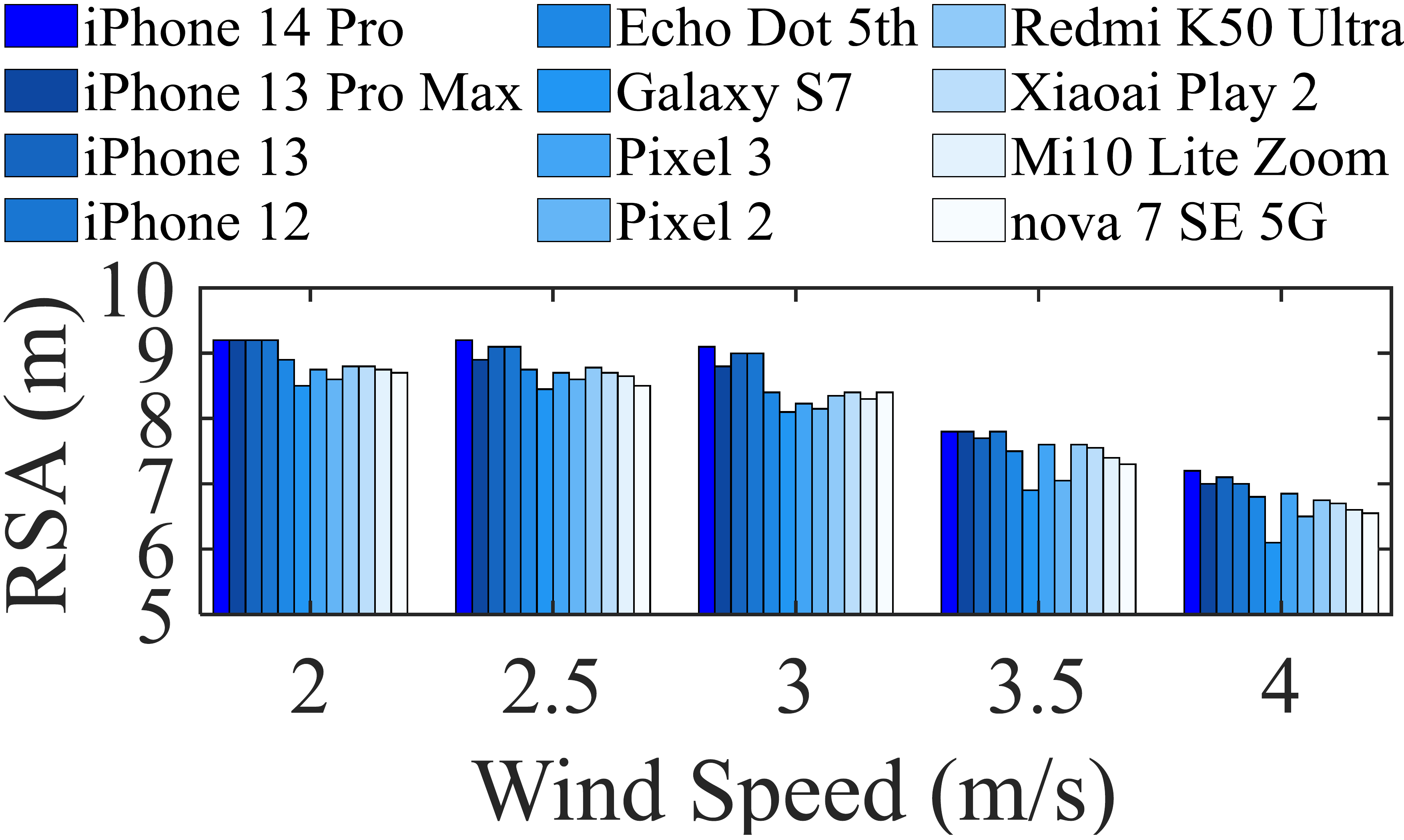}}
        \caption{Impact of wind speeds.}
        \label{wind}
    \end{minipage}
     \vspace{-5pt}
\end{figure}
\subsubsection{S1 - Long-range attacks between high-rise buildings\label{S-1}} As shown in Fig. \ref{S1}, the experiment in Scenario S1 was conducted between two high-rise buildings, taking into account high-altitude wind interference. The attacker was located in a public corridor with no power supply and frequent pedestrian traffic, making it impractical to use large equipment. The experimental results (Fig. \ref{wind}) show that \SystemName maintained a stable attack range of 8.6 meters at wind speeds of 2–3 m/s, while wind speeds above 3.5 m/s slightly affected the attack distance. However, attackers can rely on weather forecasts to avoid unfavorable wind conditions and ensure reliable long-range attacks.

\subsubsection{S2 - Attack against outdoor targets\label{S-2}} 
Due to its portability, \SystemName is well-suited for conducting covert attacks on moving targets in outdoor environments. To validate this, we recruited volunteers carrying different types of devices in the real-world scenario shown in Fig. \ref{S2}. The volunteers walked at speeds of 1 - 1.5 m/s, corresponding to normal walking and brisk walking, respectively, with the devices either held in hand or placed in pockets.

\begin{figure}[t!]
\begin{minipage}[t]{0.475\linewidth}
\setlength{\abovecaptionskip}{-0.3cm}
\centering        
\includegraphics[width=1\linewidth]{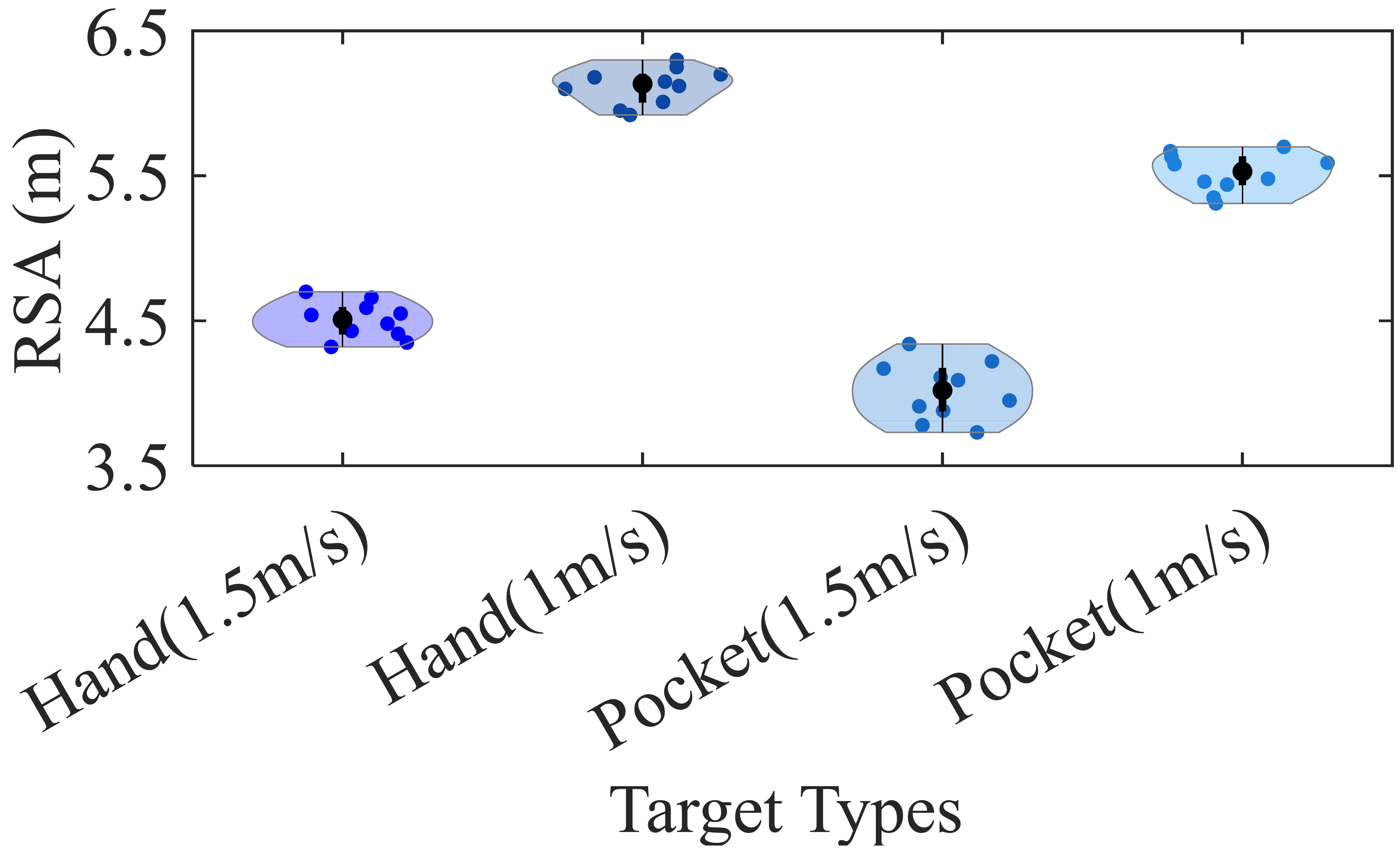}
\caption{Impact of outdoor target.}
\label{moving target}
\end{minipage}
\hfill
\begin{minipage}[t]{0.47\linewidth}
\setlength{\abovecaptionskip}{-0.3cm}
\centering
\includegraphics[width=1\linewidth]{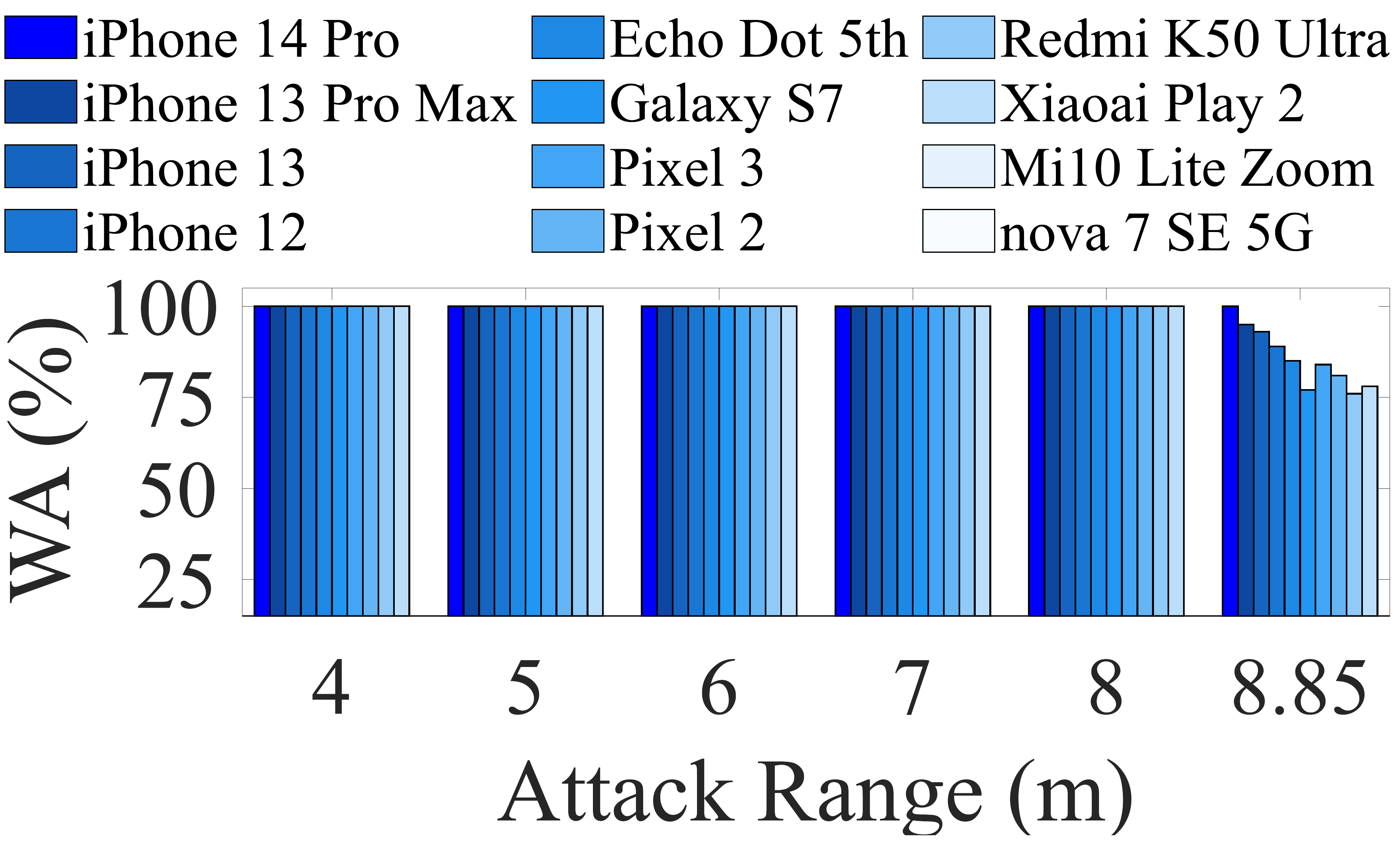}
\caption{System’s Word Acc. (WA)}
\label{accuracy}
\end{minipage}
\vspace{-5pt}
\end{figure}

The experimental results in Fig. \ref{moving target} show that when the target moves at a speed of 1 m/s and holds the device, \SystemName achieves its optimal average RSA of 6.2 meters. As the target's speed increases or the device is placed in a pocket, the attack success rate decreases. This is due to slight shaking of the device during movement, making it difficult for the system to align, and physical obstructions like a pocket weakening the signal. However, even in the harshest conditions (1.5 m/s speed with the device in the pocket), \SystemName still achieves an average RSA of 4.1 meters. As mentioned in Section \ref{Threat Model2.2}, a distance beyond 3 meters is considered a safe range unlikely to alert the target. Therefore, we believe that \SystemName can perform attacks without alerting the target. To validate the feasibility of \SystemName, we released an attack video targeting walking individuals \cite{OutdoorMetaAttack}.

\subsubsection{S3 - Eavesdropping attacks on sensitive
information \label{S-3}} Eavesdropping is a common method of information theft in commercial activities, where attackers send inaudible commands to manipulate the target phone to dial a preset number, as shown in Fig.~\ref{S3}. The main challenge in this scenario is improving character recognition accuracy. We evaluated the performance of \SystemName in transmitting the command \textit{call 1234567890} at various distances. Results show that \SystemName achieved 100\% recognition accuracy at a distance of 8.85 meters on an Apple iPhone 14 Pro, and similar accuracy on other devices within an average range of 8 meters. This demonstrates that the metamaterial-based beamforming design of \SystemName enables efficient and precise signal transmission.

\subsubsection{S4 - Effect of passing individuals on
attack range\label{S-4}} 
\begin{figure}[t!]
 \setlength{\abovecaptionskip}{-10pt}
         \begin{minipage}[t]{0.29\linewidth}
        \centering
\includegraphics[width=1\linewidth]{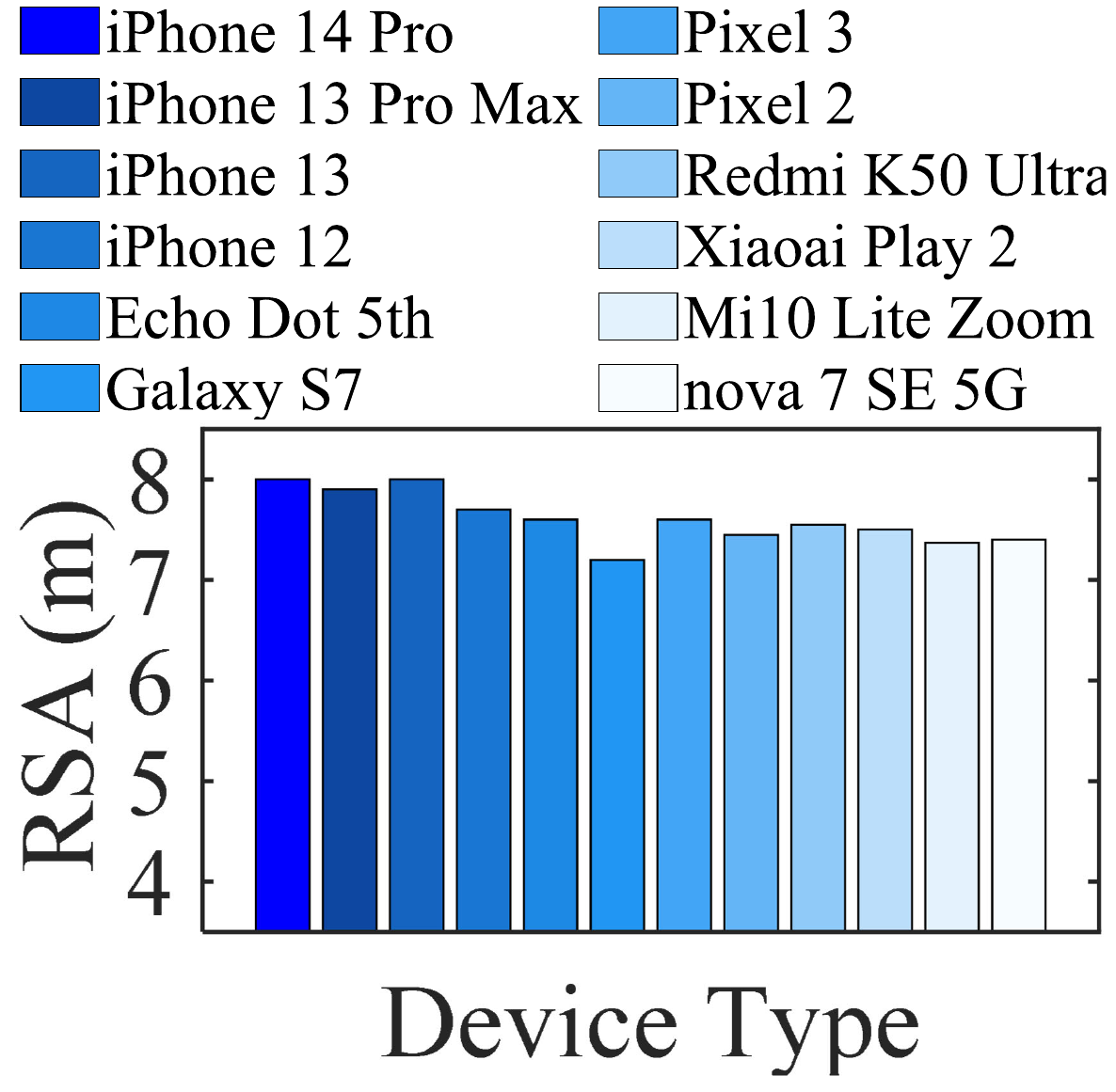}
        \caption{Effect by passing people}
        \label{fig:settingResult}
    \end{minipage}
    \hspace{0.01\textwidth}
\begin{minipage}[t]{0.69\linewidth}
        \centering               {\includegraphics[width=1\linewidth]{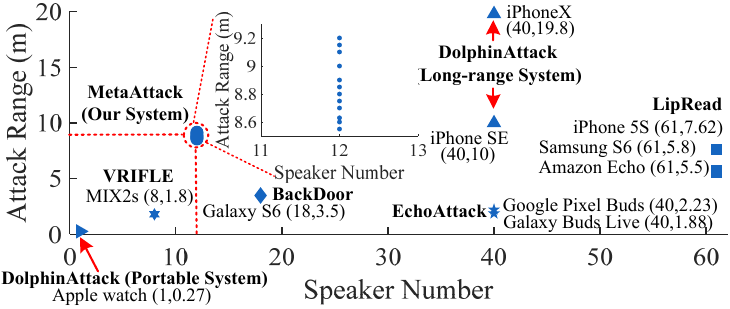}}
        \caption{Performance compared to prior research (attack range vs. number of speakers)}
        \label{Ratio}
    \end{minipage}

 \end{figure}

Passing pedestrians may affect the propagation of inaudible sound waves. To assess their impact on the attack range of \SystemName, we conducted experiments in a public study room (parameters shown in Fig. \ref{S4}). As shown in Fig. \ref{fig:settingResult}, when a person walks back and forth along the attack path at a speed of 1 m/s (using the iPhone 14 Pro as an example), the attack distance still reaches 7.6 meters with almost no performance degradation. Although pedestrians may absorb and scatter part of the ultrasonic signal, the beamforming design of \SystemName ensures that the sound waves remain highly directional and focused, allowing the signal to be transmitted effectively without being significantly dispersed, thus maintaining the effectiveness of the attack.

\begin{figure}[t!]
    \centering   
     \setlength{\abovecaptionskip}{-10pt}
    \includegraphics[width=1\linewidth]{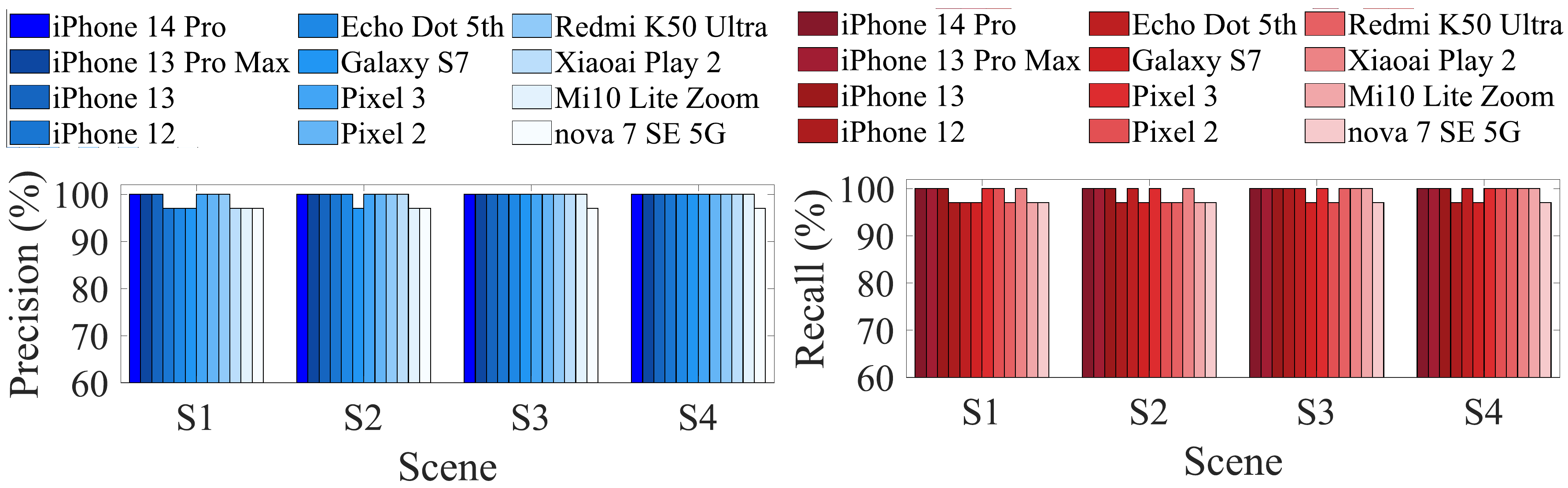}
    \caption{The feedback accuracy of \SystemName for different scenarios}
    \label{feedback}
    \vspace{-5pt}
 \end{figure}
 
\subsubsection{S5 - Feedback mechanism performance\label{S5}} The feedback mechanism is crucial for inaudible attacks, as it enables the attacker to swiftly clean the attack site to evade detection by the target. In this experiment, we evaluated the efficacy of \SystemName's feedback across scenarios S1 to S4. As shown in Fig.~\ref{feedback}, although the Precision and Recall vary slightly across different devices, they consistently remain above 97\%, demonstrating high feedback accuracy. Notably, \SystemName performs relatively worse in unstable scenarios (Fig.~\ref{S1} and Fig.~\ref{S2}), indicating that the feedback mechanism is more susceptible to external environmental interference. In such instances, the attacker must exercise timely judgment based on personal experience to avoid detection by the target, particularly in the event of a successful and continued attack.

\subsection{Performance Compared to Prior Research}\label{N3}

We made a comparison between the \SystemName and existing primary attack devices. Theoretically, increasing the number of loudspeakers can potentially increase the attack range; however, too many loudspeakers can compromise concealment. In Fig. \ref{Ratio} we show the correlation between the attack range and the number of loudspeakers. An important improvement of the \SystemName over previous studies is its invisibility and portability. As shown in Fig. \ref{Ratio}, the \SystemName uses only 12 speakers for its attacks and relies on only two 70 mm $\times$ 130 mm lithium batteries for power.  
LipRead~\cite{lipread} used 61 speakers for its operations and relied on power from a DC power supply. 
Another system called EchoAttack~\cite{r9} has a limited attack range of 2.23m with 40 speakers. 

The enhanced DolphinAttack\cite{longdol} has an extended attack range of 19.8 m, targeting the iPhone X. These enhancements are achieved through the use of 40 speakers and external accessories. However, this setup may affect the feasibility of performing covert attacks.
These results show that \SystemName is superior in terms of portability and execution of covert attacks, in contrast to previous research.
\section{DISCUSSION} \label{chap:9}

\cparagraph{Hardware devices for inaudible attacks} Exploring the impact of different attack devices on the functionality of the \SystemName system is an interesting direction for future research. By experimenting with various speaker and amplifier models, the \SystemName system is expected to improve attack range efficiency while optimizing resource consumption.

\cparagraph{Attack through walls, windows, or curtains} The attenuation of ultrasonic signals when passing through solid objects poses a significant challenge for transmitting inaudible commands at 40 kHz. This makes it difficult for \SystemName to effectively attack targets in enclosed environments. Future work will focus on addressing this issue.

\cparagraph{Audible leakage in the air propagation process} The audible leakage during airborne transmission caused by nonlinear acoustics remains an issue to be addressed. Although \SystemName has compressed the audible leakage into a narrow range of about 12°, where it can only be detected if the target's ear is within this range, the problem is not yet fully resolved. We plan to thoroughly address this limitation in future work.

\cparagraph{Optimization of Metamaterials}
Currently, \SystemName has achieved the longest possible attack range under its existing structure. We believe that further extending the attack distance hinges on enhancing the filtering performance of the metamaterials and improving their beamforming capability at lower carrier frequencies (e.g., around 25 kHz). Compared to high-frequency ultrasound, low-frequency ultrasound maintains inaudibility while offering better propagation characteristics, making it more promising for long-range transmission. In addition, due to the constraints of the fixed LAM filtering structure, the current optimization of speaker parameters (such as quantity, arrangement, and carrier frequency) may only yield locally optimal results. In future work, we plan to explore more advanced metamaterial designs and jointly optimize them with speaker parameters to enable more stable and longer-range attacks.

\cparagraph{Possible conflicts in the feedback mechanism} Considering that a few attack commands (such as "turn on airplane mode") may conflict with feedback mechanism, we can first confirm the position of the target device through feedback mechanism before executing the attack.

\section{COUNTERMEASURES} \label{chap:10}

\cparagraph{Previous defense strategies} Defense strategies against inaudible attacks fall into two main categories: forensics-based and cancellation-based. Forensics-based defenses detect attacks by analyzing signal characteristics through methods like frequency analysis or deep learning. For example, EarArray~\cite{eararray} analyzes timing differences across microphones; Li et al.~\cite{li2021robust} extract spatial features and apply deep learning to distinguish human speech from machine-generated signals; NormDetect~\cite{normal} identifies missing features in attack audio; and LipRead~\cite{lipread} analyzes power, autocorrelation, and low-frequency amplitude deviations. Cancellation-based defenses use ultrasonic speaker arrays to emit sound waves that neutralize attack signals \cite{r5}.

\cparagraph{Challenges and future directions in defending against \SystemName attacks} Forensics-based defenses are sensitive to microphone variations, leading to unstable results \cite{r5}. We replicated LipRead~\cite{lipread} in a quiet (40~dB) lab and tested 30 rounds on various devices. iPhone 13 Pro Max, iPhone 13, iPhone 12, Galaxy S7, and nova 7 SE 5G showed over 90\% success, while iPhone 14 Pro, Redmi K50 Ultra, Pixel 3, and Pixel 2 were below 63.3\%, confirming poor cross-device consistency. These defenses also disable voice assistants only after detecting attacks, disrupting normal use. Cancellation-based methods are more effective but rely on bulky equipment and have limited range, making them impractical for mobile attacks like \SystemName. Future defenses should focus on lightweight, externally integrated mechanisms that block attack signals before they reach the microphone.

\section{RELATED WORK} \label{chap:11}\cparagraph{Attacks on Voice Control Systems} Attacks on voice control systems have been proven to pose serious security threats. Existing attack methods mainly fall into the following categories: adversarial attacks, magnetic injection attacks, laser attacks, and inaudible attacks.

Adversarial attacks deceive devices by embedding carefully crafted instructions in normal audio to induce them to perform actions preset by the attacker \cite{CommanderSong,Metamorph,Robust,Practical,kenku,ALIF,qfa2sr,devil,smack}. For example, CommanderSong \cite{CommanderSong} hides commands within songs and performs the attack by playing the song. Metamorph \cite{Metamorph} enhances robustness using pre-coding, extending the attack range to 6 meters. In addition, MagBackdoor \cite{Magbackdoor} injects command signals into voice control systems through the device’s charging port, while Sugawara et al. \cite{r97} use long-range lasers to simulate sound vibrations to attack voice control systems. Inaudible attacks can silently interfere with or control voice control systems, making them more covert \cite{r7,dol,r9,lipread,nuit,longdol,perturbation,surfingattack}. For example, DolphinAttack \cite{dol} achieves control over smart devices by modulating attack commands onto ultrasound waves. LipRead \cite{r9} and DolphinAttack (Long-range System) \cite{longdol} further extend the attack range by incorporating spectrum splitting techniques and large speaker arrays. VRIFLE \cite{perturbation} injects ultrasonic signals while the target is speaking, interfering with the target's control of the device, thereby increasing the destructiveness of the attack. \SystemName first combines attacks with acoustic metamaterials to achieve long-range attacks in a compact size, offering both concealment and threat.

\cparagraph{Acoustic Metamaterials} Acoustic metamaterials designed for ventilation barriers are an emerging research area, enabling both filtering and airflow in a compact form. For example, Ghaffarivardavagh et al. \cite{r31} proposed a deep subwavelength acoustic metamaterial based on Fano-like resonance, and Zhu et al. \cite{r14} expanded its filtering bandwidth using nonlocal effects. Building on this, \SystemName further extends the filtering range to 100 – 4000`\textit{Hz} and pioneers its application in ultrasonic attacks.

\section{CONCLUSION} \label{chap:12}
This paper has presented the first inaudible attack based on acoustical materials. This new attack significantly extends the attack range while maintaining a compact setup. We have enhanced conventional acoustical materials in two ways. Firstly, we optimized the structure of the acoustic metamaterial to filter out audible leakage resulting from inaudible attacks effectively. Secondly, we transformed the acoustic metamaterial into a piston sound source, enabling the use of a small speaker array to achieve the same beamforming effect as larger devices. Importantly, our methodology utilizes commercially available off-the-shelf components, ensuring practical implementation.  We tested the performance of our design under various experiments against prior methods. Our approach can achieve covert and longer-range inaudible attacks using a smallest device compared to prior works.

\section{Ethical Considerations} \label{chap:13}

\cparagraph{Ethical Challenges in Reproducibility and Disclosure} Although publishing the 3D model designs of \SystemName would aid in the reproducibility of the experiment, it could also provide malicious actors with the information necessary to execute attacks. Therefore, we have decided to share the related materials only with verified researchers who contact us, balancing transparency with security.

\cparagraph{Ethical Considerations in Experiments and Testing} The research equipment was purchased at our own expense, and all experiments were approved by the university's Institutional Review Board (IRB), ensuring compliance with data security and ethical standards. All participants voluntarily agreed to participate in the experiments based on informed consent, ensuring adherence to ethical requirements.
\bibliography{references}

\begin{thebibliography}{10}
\providecommand{\url}[1]{#1}
\csname url@samestyle\endcsname
\providecommand{\newblock}{\relax}
\providecommand{\bibinfo}[2]{#2}
\providecommand{\BIBentrySTDinterwordspacing}{\spaceskip=0pt\relax}
\providecommand{\BIBentryALTinterwordstretchfactor}{4}
\providecommand{\BIBentryALTinterwordspacing}{\spaceskip=\fontdimen2\font plus
\BIBentryALTinterwordstretchfactor\fontdimen3\font minus \fontdimen4\font\relax}
\providecommand{\BIBforeignlanguage}[2]{{%
\expandafter\ifx\csname l@#1\endcsname\relax
\typeout{** WARNING: IEEEtran.bst: No hyphenation pattern has been}%
\typeout{** loaded for the language `#1'. Using the pattern for}%
\typeout{** the default language instead.}%
\else
\language=\csname l@#1\endcsname
\fi
#2}}
\providecommand{\BIBdecl}{\relax}
\BIBdecl

\bibitem{voicecloak}
M.~Chen, L.~Lu, J.~Wang, J.~Yu, Y.~Chen, Z.~Wang, Z.~Ba, F.~Lin, and K.~Ren, ``Voicecloak: Adversarial example enabled voice de-identification with balanced privacy and utility,'' \emph{Proceedings of the ACM on Interactive, Mobile, Wearable and Ubiquitous Technologies}, vol.~7, no.~2, pp. 1--21, 2023.

\bibitem{hu2022accear}
P.~Hu, H.~Zhuang, P.~S. Santhalingam, R.~Spolaor, P.~Pathak, G.~Zhang, and X.~Cheng, ``Accear: Accelerometer acoustic eavesdropping with unconstrained vocabulary,'' in \emph{2022 IEEE Symposium on Security and Privacy (SP)}.\hskip 1em plus 0.5em minus 0.4em\relax IEEE, 2022, pp. 1757--1773.

\bibitem{hu2022milliear}
P.~Hu, Y.~Ma, P.~S. Santhalingam, P.~H. Pathak, and X.~Cheng, ``Milliear: Millimeter-wave acoustic eavesdropping with unconstrained vocabulary,'' in \emph{IEEE INFOCOM 2022-IEEE Conference on Computer Communications}.\hskip 1em plus 0.5em minus 0.4em\relax IEEE, 2022, pp. 11--20.

\bibitem{hu2022towards}
P.~Hu, W.~Li, Y.~Ma, P.~S. Santhalingam, P.~Pathak, H.~Li, H.~Zhang, G.~Zhang, X.~Cheng, and P.~Mohapatra, ``Towards unconstrained vocabulary eavesdropping with mmwave radar using gan,'' \emph{IEEE Transactions on Mobile Computing}, vol.~23, no.~1, pp. 941--954, 2022.

\bibitem{hu2023mmecho}
P.~Hu, W.~Li, R.~Spolaor, and X.~Cheng, ``mmecho: A mmwave-based acoustic eavesdropping method,'' in \emph{2023 IEEE Symposium on Security and Privacy (S\&P)}.\hskip 1em plus 0.5em minus 0.4em\relax https://doi.org/10.1109/SP46215.2023.10179484, 2023, pp. 1840--1856.

\bibitem{zhang2024echolight}
G.~Zhang, Z.~Xiang, H.~Fu, Y.~Yang, and P.~Hu, ``Echolight: Sound eavesdropping based on ambient light reflection,'' in \emph{IEEE INFOCOM 2024-IEEE Conference on Computer Communications}.\hskip 1em plus 0.5em minus 0.4em\relax IEEE, 2024, pp. 341--350.

\bibitem{yang2024rf}
Y.~Yang, G.~Wang, Z.~An, G.~Zhang, X.~Cheng, and P.~Hu, ``Rf-parrot: Wireless eavesdropping on wired audio,'' in \emph{IEEE INFOCOM 2024-IEEE Conference on Computer Communications}.\hskip 1em plus 0.5em minus 0.4em\relax IEEE, 2024, pp. 701--710.

\bibitem{wang2024wireless}
G.~Wang, Z.~Shi, Y.~Yang, Z.~An, G.~Zhang, P.~Hu, X.~Cheng, and J.~Cao, ``Wireless eavesdropping on wired audio with radio-frequency retroreflector attack,'' \emph{IEEE Transactions on Mobile Computing}, 2024.

\bibitem{handling}
T.~Yu, V.~Sekar, S.~Seshan, Y.~Agarwal, and C.~Xu, ``Handling a trillion (unfixable) flaws on a billion devices: Rethinking network security for the internet-of-things,'' in \emph{Proceedings of the 14th ACM workshop on hot topics in networks}, 2015, pp. 1--7.

\bibitem{watching}
L.~Wang, X.~Zhang, Y.~Jiang, Y.~Zhang, C.~Xu, R.~Gao, and D.~Zhang, ``Watching your phone's back: Gesture recognition by sensing acoustical structure-borne propagation,'' \emph{Proceedings of the ACM on Interactive, Mobile, Wearable and Ubiquitous Technologies}, vol.~5, no.~2, pp. 1--26, 2021.

\bibitem{r7}
N.~Roy, H.~Hassanieh, and R.~Roy~Choudhury, ``Backdoor: Making microphones hear inaudible sounds,'' in \emph{Proceedings of the 15th Annual International Conference on Mobile Systems, Applications, and Services}, 2017, pp. 2--14.

\bibitem{dol}
G.~Zhang, C.~Yan, X.~Ji, T.~Zhang, T.~Zhang, and W.~Xu, ``Dolphinattack: Inaudible voice commands,'' in \emph{Proceedings of the 2017 ACM SIGSAC conference on computer and communications security}, 2017, pp. 103--117.

\bibitem{r9}
G.~Li, Z.~Cao, and T.~Li, ``Echoattack: Practical inaudible attacks to smart earbuds,'' in \emph{Proceedings of the 21st Annual International Conference on Mobile Systems, Applications and Services}, 2023, pp. 383--396.

\bibitem{r100}
Q.~Wang, P.~Guo, and L.~Xie, ``Inaudible adversarial perturbations for targeted attack in speaker recognition,'' \emph{arXiv preprint arXiv:2005.10637}, 2020.

\bibitem{r101}
Y.~Wang, H.~Guo, and Q.~Yan, ``Ghosttalk: Interactive attack on smartphone voice system through power line,'' \emph{arXiv preprint arXiv:2202.02585}, 2022.

\bibitem{Fencesitter}
J.~Deng, Y.~Chen, and W.~Xu, ``Fencesitter: Black-box, content-agnostic, and synchronization-free enrollment-phase attacks on speaker recognition systems,'' in \emph{Proceedings of the 2022 ACM SIGSAC Conference on Computer and Communications Security}, 2022, pp. 755--767.

\bibitem{Magbackdoor}
T.~Liu, F.~Lin, Z.~Wang, C.~Wang, Z.~Ba, L.~Lu, W.~Xu, and K.~Ren, ``Magbackdoor: Beware of your loudspeaker as a backdoor for magnetic injection attacks,'' in \emph{2023 IEEE Symposium on Security and Privacy (SP)}.\hskip 1em plus 0.5em minus 0.4em\relax IEEE, 2023, pp. 3416--3431.

\bibitem{perturbation}
X.~Li, C.~Yan, X.~Lu, Z.~Zeng, X.~Ji, and W.~Xu, ``Inaudible adversarial perturbation: Manipulating the recognition of user speech in real time,'' \emph{arXiv preprint arXiv:2308.01040}, 2023.

\bibitem{lipread}
N.~Roy, S.~Shen, H.~Hassanieh, and R.~R. Choudhury, ``Inaudible voice commands: The $\{$Long-Range$\}$ attack and defense,'' in \emph{15th USENIX Symposium on Networked Systems Design and Implementation (NSDI 18)}, 2018, pp. 547--560.

\bibitem{longdol}
C.~Yan, G.~Zhang, X.~Ji, T.~Zhang, T.~Zhang, and W.~Xu, ``The feasibility of injecting inaudible voice commands to voice assistants,'' \emph{IEEE Transactions on Dependable and Secure Computing}, vol.~18, no.~3, pp. 1108--1124, 2019.

\bibitem{r75}
M.~Xu, W.~S. Harley, Z.~Ma, P.~V. Lee, and D.~J. Collins, ``Sound-speed modifying acoustic metasurfaces for acoustic holography,'' \emph{Advanced Materials}, vol.~35, no.~14, p. 2208002, 2023.

\bibitem{r76}
X.~Wang, R.~Dong, Y.~Li, and Y.~Jing, ``Non-local and non-hermitian acoustic metasurfaces,'' \emph{Reports on Progress in Physics}, 2023.

\bibitem{r77}
J.~Guo, R.~Qu, Y.~Fang, W.~Yi, and X.~Zhang, ``A phase-gradient acoustic metasurface for broadband duct noise attenuation in the presence of flow,'' \emph{International Journal of Mechanical Sciences}, vol. 237, p. 107822, 2023.

\bibitem{r78}
F.~Mir, D.~Mandal, and S.~Banerjee, ``Metamaterials for acoustic noise filtering and energy harvesting,'' \emph{Sensors}, vol.~23, no.~9, p. 4227, 2023.

\bibitem{r79}
X.~Song, T.~Chen, W.~Huang, and C.~Chen, ``Frequency-selective modulation of reflected wave fronts using a four-mode coding acoustic metasurface,'' \emph{Physics Letters A}, vol. 394, p. 127145, 2021.

\bibitem{r106}
S.~A. Cummer, J.~Christensen, and A.~Al{\`u}, ``Controlling sound with acoustic metamaterials,'' \emph{Nature Reviews Materials}, vol.~1, no.~3, pp. 1--13, 2016.

\bibitem{r11}
B.~Assouar, B.~Liang, Y.~Wu, Y.~Li, J.-C. Cheng, and Y.~Jing, ``Acoustic metasurfaces,'' \emph{Nature Reviews Materials}, vol.~3, no.~12, pp. 460--472, 2018.

\bibitem{r12}
G.~Ma and P.~Sheng, ``Acoustic metamaterials: From local resonances to broad horizons,'' \emph{Science advances}, vol.~2, no.~2, p. e1501595, 2016.

\bibitem{r13}
G.~Memoli, L.~Chisari, J.~P. Eccles, M.~Caleap, B.~W. Drinkwater, and S.~Subramanian, ``Vari-sound: A varifocal lens for sound,'' in \emph{Proceedings of the 2019 CHI Conference on Human Factors in Computing Systems}, 2019, pp. 1--14.

\bibitem{r14}
Y.~Zhu, R.~Dong, D.~Mao, X.~Wang, and Y.~Li, ``Nonlocal ventilating metasurfaces,'' \emph{Physical Review Applied}, vol.~19, no.~1, p. 014067, 2023.

\bibitem{r15}
Z.~Ren, Y.~Cheng, M.~Chen, X.~Yuan, and D.~Fang, ``A compact multifunctional metastructure for low-frequency broadband sound absorption and crash energy dissipation,'' \emph{Materials \& Design}, vol. 215, p. 110462, 2022.

\bibitem{r16}
Z.~Zhou, S.~Huang, D.~Li, J.~Zhu, and Y.~Li, ``Broadband impedance modulation via non-local acoustic metamaterials,'' \emph{National Science Review}, vol.~9, no.~8, p. nwab171, 2022.

\bibitem{r17}
Y.-F. Zhu, A.~Merkel, K.~Donda, S.~Fan, L.~Cao, and B.~Assouar, ``Nonlocal acoustic metasurface for ultrabroadband sound absorption,'' \emph{Physical Review B}, vol. 103, no.~6, p. 064102, 2021.

\bibitem{r86}
A.~Williams~Jr, ``The piston source at high frequencies,'' \emph{The Journal of the Acoustical Society of America}, vol.~23, no.~1, pp. 1--6, 1951.

\bibitem{r118}
V.~Khokhlova, R.~Souchon, J.~Tavakkoli, O.~Sapozhnikov, and D.~Cathignol, ``Numerical modeling of finite-amplitude sound beams: Shock formation in the near field of a cw plane piston source,'' \emph{The Journal of the Acoustical Society of America}, vol. 110, no.~1, pp. 95--108, 2001.

\bibitem{MetaAttack}
``Process demonstration of metaattack,'' \url{https://youtu.be/xBlKmGy-XJc}, last accessed: 2024-9-4.

\bibitem{OutdoorMetaAttack}
``Demonstration of using metaattack in outdoor scene,'' \url{https://youtu.be/Tjw-weqvJBw}, last accessed: 2024-9-4.

\bibitem{feedback}
``Feedback code,'' \url{https://github.com/MetaAttack/MetaAttack}, last accessed: 2024-9-4.

\bibitem{r99}
V.~P. Richmond, ``Nonverbal behavior in interpersonal relations,'' \emph{(No Title)}, p. 366, 2008.

\bibitem{privacy2}
E.~Sundstrom and I.~Altman, ``Interpersonal relationships and personal space: Research review and theoretical model,'' \emph{Human Ecology}, vol.~4, pp. 47--67, 1976.

\bibitem{privacy3}
M.~J. Wieser, P.~Pauli, M.~Grosseibl, I.~Molzow, and A.~M{\"u}hlberger, ``Virtual social interactions in social anxiety—the impact of sex, gaze, and interpersonal distance,'' \emph{Cyberpsychology, Behavior, and Social Networking}, vol.~13, no.~5, pp. 547--554, 2010.

\bibitem{nuit}
Q.~Xia, Q.~Chen, and S.~Xu, ``$\{$Near-Ultrasound$\}$ inaudible trojan (nuit): Exploiting your speaker to attack your microphone,'' in \emph{32nd USENIX Security Symposium (USENIX Security 23)}, 2023, pp. 4589--4606.

\bibitem{r112}
C.~Goffaux, J.~S{\'a}nchez-Dehesa, A.~L. Yeyati, P.~Lambin, A.~Khelif, J.~Vasseur, and B.~Djafari-Rouhani, ``Evidence of fano-like interference phenomena in locally resonant materials,'' \emph{Physical review letters}, vol.~88, no.~22, p. 225502, 2002.

\bibitem{r113}
J.~A. Fan, K.~Bao, C.~Wu, J.~Bao, R.~Bardhan, N.~J. Halas, V.~N. Manoharan, G.~Shvets, P.~Nordlander, and F.~Capasso, ``Fano-like interference in self-assembled plasmonic quadrumer clusters,'' \emph{Nano letters}, vol.~10, no.~11, pp. 4680--4685, 2010.

\bibitem{Google}
``Google text-to-speech ai,'' \url{https://cloud.google.com/speech-to-text}, last accessed: 2024-8-6.

\bibitem{surfingattack}
Q.~Yan, K.~Liu, Q.~Zhou, H.~Guo, and N.~Zhang, ``Surfingattack: Interactive hidden attack on voice assistants using ultrasonic guided waves,'' in \emph{Network and Distributed Systems Security (NDSS) Symposium}, 2020.

\bibitem{delete1}
J.~Zdziarski, ``Identifying back doors, attack points, and surveillance mechanisms in ios devices,'' \emph{Digital Investigation}, vol.~11, no.~1, pp. 3--19, 2014.

\bibitem{delete2}
J.~Bellizzi, E.~Losiouk, M.~Conti, C.~Colombo, and M.~Vella, ``Vedrando: A novel way to reveal stealthy attack steps on android through memory forensics,'' \emph{Journal of Cybersecurity and Privacy}, vol.~3, no.~3, pp. 364--395, 2023.

\bibitem{delete3}
N.~Ullah, M.~Zahra, B.~Saleem, M.~Haseeb, M.~Mughal, and Z.~Muhammad, ``Delsec: An anti-forensics data deletion framework for smartphones, iot, and edge devices,'' in \emph{2024 International Conference on Engineering \& Computing Technologies (ICECT)}.\hskip 1em plus 0.5em minus 0.4em\relax IEEE, 2024, pp. 1--6.

\bibitem{iwlist}
``iwlist,'' \url{https://manpages.ubuntu.com/manpages/focal/man8/iwlist.8.html}, last accessed: 2024-8-6.

\bibitem{r31}
R.~Ghaffarivardavagh, J.~Nikolajczyk, S.~Anderson, and X.~Zhang, ``Ultra-open acoustic metamaterial silencer based on fano-like interference,'' \emph{Physical Review B}, vol.~99, no.~2, p. 024302, 2019.

\bibitem{r32}
R.~Dong, D.~Mao, Y.~Zhu, F.~Mo, X.~Wang, and Y.~Li, ``A ventilating acoustic barrier for attenuating broadband diffuse sound,'' \emph{Applied Physics Letters}, vol. 119, no.~26, 2021.

\bibitem{r33}
M.~Sun, X.~Fang, D.~Mao, X.~Wang, and Y.~Li, ``Broadband acoustic ventilation barriers,'' \emph{Physical Review Applied}, vol.~13, no.~4, p. 044028, 2020.

\bibitem{r95}
R.~Dong, M.~Sun, F.~Mo, D.~Mao, X.~Wang, and Y.~Li, ``Recent advances in acoustic ventilation barriers,'' \emph{Journal of Physics D: Applied Physics}, vol.~54, no.~40, p. 403002, 2021.

\bibitem{r80}
T.~Chen, C.~Wang, and D.~Yu, ``Pressure amplification and directional acoustic sensing based on a gradient metamaterial coupled with space-coiling structure,'' \emph{Mechanical Systems and Signal Processing}, vol. 181, p. 109499, 2022.

\bibitem{r81}
A.~De, J.~L. Drobitch, S.~Majumder, S.~Barman, S.~Bandyopadhyay, and A.~Barman, ``Resonant amplification of intrinsic magnon modes and generation of new extrinsic modes in a two-dimensional array of interacting multiferroic nanomagnets by surface acoustic waves,'' \emph{Nanoscale}, vol.~13, no.~22, pp. 10\,016--10\,023, 2021.

\bibitem{Applesiri}
``Apple siri,'' \url{https://www.apple.com/siri/}, last accessed: 2024-8-14.

\bibitem{AmazonAlexa}
``Amazon alexa,'' \url{https://alexa.amazon.com/}, last accessed: 2024-8-14.

\bibitem{GoogleAssistant}
``Google assistant,'' \url{https://assistant.google.com/}, last accessed: 2024-8-14.

\bibitem{XiaomiXiaoai}
``Xiaomi xiaoai,'' \url{https://xiaoai.mi.com/}, last accessed: 2024-8-14.

\bibitem{HuaweiXiaoyi}
``Huawei xiaoyi,'' \url{https://www.nanzhao.site/ux-work/huawei-xiaoyi}, last accessed: 2024-8-14.

\bibitem{eararray}
G.~Zhang, X.~Ji, X.~Li, G.~Qu, and W.~Xu, ``Eararray: Defending against dolphinattack via acoustic attenuation.'' in \emph{NDSS}, 2021.

\bibitem{normal}
X.~Li, X.~Ji, C.~Yan, C.~Li, Y.~Li, Z.~Zhang, and W.~Xu, ``Learning normality is enough: a software-based mitigation against inaudible voice attacks,'' in \emph{32nd USENIX Security Symposium (USENIX Security 23)}, 2023, pp. 2455--2472.

\bibitem{2014inaudible}
L.~Deshotels, ``Inaudible sound as a covert channel in mobile devices,'' in \emph{8th USENIX Workshop on Offensive Technologies (WOOT 14)}, 2014.

\bibitem{li2021robust}
Z.~Li, C.~Shi, T.~Zhang, Y.~Xie, J.~Liu, B.~Yuan, and Y.~Chen, ``Robust detection of machine-induced audio attacks in intelligent audio systems with microphone array,'' in \emph{Proceedings of the 2021 ACM SIGSAC Conference on Computer and Communications Security}, 2021, pp. 1884--1899.

\bibitem{r5}
Y.~He, J.~Bian, X.~Tong, Z.~Qian, W.~Zhu, X.~Tian, and X.~Wang, ``Canceling inaudible voice commands against voice control systems,'' in \emph{The 25th Annual International Conference on Mobile Computing and Networking}, 2019, pp. 1--15.

\bibitem{CommanderSong}
X.~Yuan, Y.~Chen, Y.~Zhao, Y.~Long, X.~Liu, K.~Chen, S.~Zhang, H.~Huang, X.~Wang, and C.~A. Gunter, ``$\{$CommanderSong$\}$: A systematic approach for practical adversarial voice recognition,'' in \emph{27th USENIX security symposium (USENIX security 18)}, 2018, pp. 49--64.

\bibitem{Metamorph}
T.~Chen, L.~Shangguan, Z.~Li, and K.~Jamieson, ``Metamorph: Injecting inaudible commands into over-the-air voice controlled systems,'' in \emph{Network and Distributed Systems Security (NDSS) Symposium}, 2020.

\bibitem{Robust}
H.~Yakura and J.~Sakuma, ``Robust audio adversarial example for a physical attack,'' \emph{arXiv preprint arXiv:1810.11793}, 2018.

\bibitem{Practical}
H.~Abdullah, W.~Garcia, C.~Peeters, P.~Traynor, K.~R. Butler, and J.~Wilson, ``Practical hidden voice attacks against speech and speaker recognition systems,'' \emph{arXiv preprint arXiv:1904.05734}, 2019.

\bibitem{kenku}
X.~Wu, S.~Ma, C.~Shen, C.~Lin, Q.~Wang, Q.~Li, and Y.~Rao, ``$\{$KENKU$\}$: Towards efficient and stealthy black-box adversarial attacks against $\{$ASR$\}$ systems,'' in \emph{32nd USENIX Security Symposium (USENIX Security 23)}, 2023, pp. 247--264.

\bibitem{ALIF}
P.~Cheng, Y.~Wang, P.~Huang, Z.~Ba, X.~Lin, F.~Lin, L.~Lu, and K.~Ren, ``Alif: Low-cost adversarial audio attacks on black-box speech platforms using linguistic features,'' in \emph{2024 IEEE Symposium on Security and Privacy (SP)}.\hskip 1em plus 0.5em minus 0.4em\relax IEEE, 2024, pp. 1628--1645.

\bibitem{qfa2sr}
G.~Chen, Y.~Zhang, Z.~Zhao, and F.~Song, ``$\{$QFA2SR$\}$:$\{$Query-Free$\}$ adversarial transfer attacks to speaker recognition systems,'' in \emph{32nd USENIX Security Symposium (USENIX Security 23)}, 2023, pp. 2437--2454.

\bibitem{devil}
Y.~Chen, X.~Yuan, J.~Zhang, Y.~Zhao, S.~Zhang, K.~Chen, and X.~Wang, ``$\{$Devil’s$\}$ whisper: A general approach for physical adversarial attacks against commercial black-box speech recognition devices,'' in \emph{29th USENIX Security Symposium (USENIX Security 20)}, 2020, pp. 2667--2684.

\bibitem{smack}
Z.~Yu, Y.~Chang, N.~Zhang, and C.~Xiao, ``$\{$SMACK$\}$: Semantically meaningful adversarial audio attack,'' in \emph{32nd USENIX security symposium (USENIX security 23)}, 2023, pp. 3799--3816.

\bibitem{r97}
T.~Sugawara, B.~Cyr, S.~Rampazzi, D.~Genkin, and K.~Fu, ``Light commands:$\{$Laser-Based$\}$ audio injection attacks on $\{$Voice-Controllable$\}$ systems,'' in \emph{29th USENIX Security Symposium (USENIX Security 20)}, 2020, pp. 2631--2648.

\end{thebibliography}
\bibliographystyle{IEEEtran}

\vspace{-23pt}
\begin{IEEEbiography}[{\includegraphics[width=1in,height=1.25in,clip,keepaspectratio]{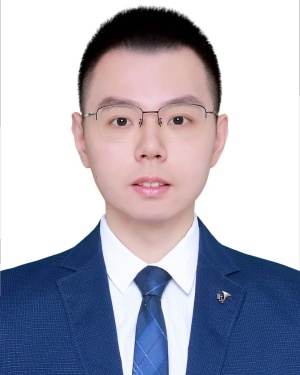}}]{Zhiyuan Ning}
is currently pursuing a Ph.D. degree at the School of Information Science and Technology, Northwest University. His research interests include acoustic metamaterials and the security of acoustic Internet of Things (IoT) systems.

\end{IEEEbiography}

\vspace{-33pt}
\begin{IEEEbiography}[{\includegraphics[width=1in,height=1.25in,clip,keepaspectratio]{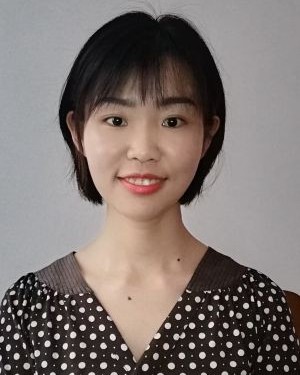}}]{Juan He}
is a PH.D student at the School of Information Science and Technology, Northwest University, and is currently pursuing a Ph.D. in Software Engineering at the same institution. Her research focuses on acoustic-based sensing, security, communication, and performance optimization within the Internet of Things.
\end{IEEEbiography}

\vspace{-33pt}
\begin{IEEEbiography}[{\includegraphics[width=1in,height=1.25in,clip,keepaspectratio]{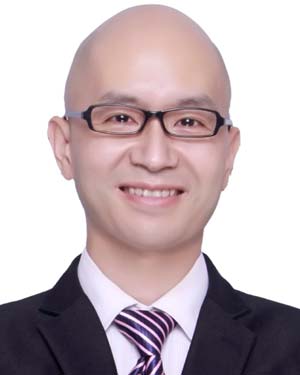}}]{Zhanyong Tang} received the Ph.D. degree in computer software and theory from Northwest University, Xi’an, China, in 2014. He is currently a professor with the School of Information Science and Technology, Northwest University. His research interests include software and system security, mobile computing, and programming languages.
\end{IEEEbiography}

\vspace{-33pt}
\begin{IEEEbiography}[{\includegraphics[width=1in,height=1.25in,clip,keepaspectratio]{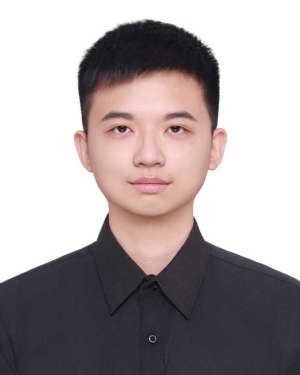}}]{Weihang Hu} has received his Master's degree from the School of Information Science and Technology, Northwest University. His primary research focus is on acoustic metamaterials.
\end{IEEEbiography}

\vspace{-33pt}
\begin{IEEEbiography}[{\includegraphics[width=1in,height=1.25in,clip,keepaspectratio]{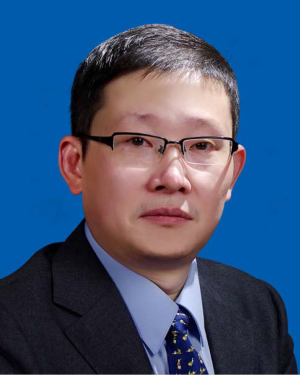}}]{Xiaojiang Chen} is a Professor with the School of Information Science and Technology, Northwest University. He received his Ph.D. degree in computer software and theory from Northwest University, Xi’an, China, in 2010.  His current research interests include RF-based sensing and performance issues in internet of things.
\end{IEEEbiography}

\vspace{11pt}

\vfill
\end{document}